\documentclass[journal]{IEEEtran}

\ifCLASSINFOpdf
\else
\fi

\usepackage{textcomp}

\usepackage{amsmath}
\usepackage{amssymb}
\usepackage{graphicx}
\usepackage{tikz}
\usepackage{multicol}
\usepackage{pifont}
\usepackage{enumerate}

\newtheorem{mylemma}{\it Lemma}
\newtheorem{mytheorem}{\it Theorem}
\newtheorem{myproposition}{\it Proposition}
\newtheorem{myproof}{\it Proof}
\newtheorem{mydefinition}{\it Definition}
\newtheorem{myexample}{\it Example}

\makeatletter
\def\footnoterule{\relax%
  \kern-5pt
  \hbox to \columnwidth{\hfill\vrule width 1\columnwidth height 0.8pt\hfill}
  \kern4.6pt}
\makeatother

\begin{document}
\title{Equidistant Polarizing Transforms}

\author{{Sinan Kahraman}
\thanks{S.~Kahraman was with the Department of Electrical and Electronics Engineering, Bilkent University,
Ankara 06800, Turkey
(e-mail: sinan.kahraman@bilkent.edu.tr).}
}

\markboth{R}%
{R}

\onecolumn

\maketitle

\begin{abstract}
We consider non-binary polarization transform problem of polar codes.
The focus of this work is on finding equidistant (or "almost" equidistant) polarizing transforms to maximize the minimum distance and to approach the equidistant distant spectrum bound as a function of signal set.
This bound is an ultimate limit and tight for additive white Gaussian channels. 
In this way, polarization of error probability for the good channel is improved while the bad channel is almost the same for a given signal set.
Our main result is that the polarization speed is increased by using polarizing transform with an improved distance profile.   
A non-binary polarization transform for $q=5$ is found that can provide the equidistant distant spectrum bound.
Almost equidistant polarization transforms for $q=4,6,7$ and $8$ are introduced for PSK signal sets. 
As a private solution, we provided new geometries for the signal set design to define the equidistant transform for $q=3$ and $q=4$ for one and two dimensional signalings.
We proposed a procedure to design transforms that have better distance profiles for q-ary signal sets.
Finally, we show improvements in frame error rates.

\end{abstract}


\IEEEpeerreviewmaketitle

\section{Introduction} 
\begin{figure}[hp]
\centering
\begin{tikzpicture}[thick,scale=0.7, every node/.style={scale=0.9}]
\draw[->]  (-1,0)  -- (-0.2,0) ;
\draw[->]  (0.7,0)  -- (1+1-0.5,0) ;
\draw[->]  (0.25,-2)  -- (0.25,-1.2) ;
\draw[->]  (-1,-2)  -- (1+1-0.5,-2) ;
\draw (-0.2,-1.2)  -- (0.7,-1.2);
\draw (-0.2,0.5)  -- (0.7,0.5);
\draw (-0.2,-1.2)  -- (-0.2,0.5);
\draw (0.7,-1.2)  -- (0.7,0.5);
\begin{scope}[thick]
\draw (0.25,-0.35) node[] {$f$};
\end{scope}
\draw (-1,0) node[left] {$u_1$};
\draw (-1,-2) node[left] {$u_2$};
\draw (1+2-1.5,0-0.03) node[right] {$x_1$};
\draw (1+2-1.5,-2-0.03) node[right] {$x_2$};
\end{tikzpicture}
\caption{A basic scheme with a polarizing transform $f$ for q-ary input alphabet.}
\label{Fig_1}
\end{figure}
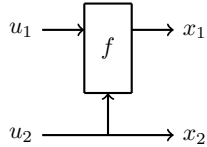

The goal of this paper is to present a method for improving the minimum distance of non-binary polar codes. Following the notation in \cite{arikan_channel_2009}, we consider a memoryless channel $W:\mathcal{X} \rightarrow \mathcal{Y}$ with input alphabet $\mathcal{X}$, output alphabet $\mathcal{Y}$, and transition probabilities $\{W(y|x):x \in \mathcal{X} ,y \in \mathcal{Y} \}$. We assume that $\mathcal{X}$ is a finite set and label its elements such that $\mathcal{X}=\{0,1,\dots,q-1\}$, where $q\geq2$ is an arbitrary integer. We leave $\mathcal{Y}$ arbitrary for the moment. We will consider polarization schemes based on a basic transform of the type depicted in Fig.~\ref{Fig_1}.

The transform is defined by a \textsl{kernel} $$f:\mathcal{X}^2 \rightarrow \mathcal{X},$$ which we assume is a mapping with the following properties:
\begin{itemize}
\item for any fixed $u_1 \in \mathcal{X}, u_2 \rightarrow f(u_1,u_2)$ is an invertible function of $u_2$;
\item for any fixed $u_2 \in \mathcal{X}, u_1 \rightarrow f(u_1,u_2)$ is an invertible function of $u_1$. 
\end{itemize}
The standard polar coding kernel as defined in \cite{arikan_channel_2009} is a mapping of this type with $f(u_1,u_2)=u_1\oplus u_2$, where $\oplus$ denotes addition mod-$q$. The present paper shows that it is possible to construct polar codes with better distance properties (hence better performance) using alternative kernels of the above type. 

\begin{figure}[!t]
\centering
\includegraphics[width=0.5\textwidth]{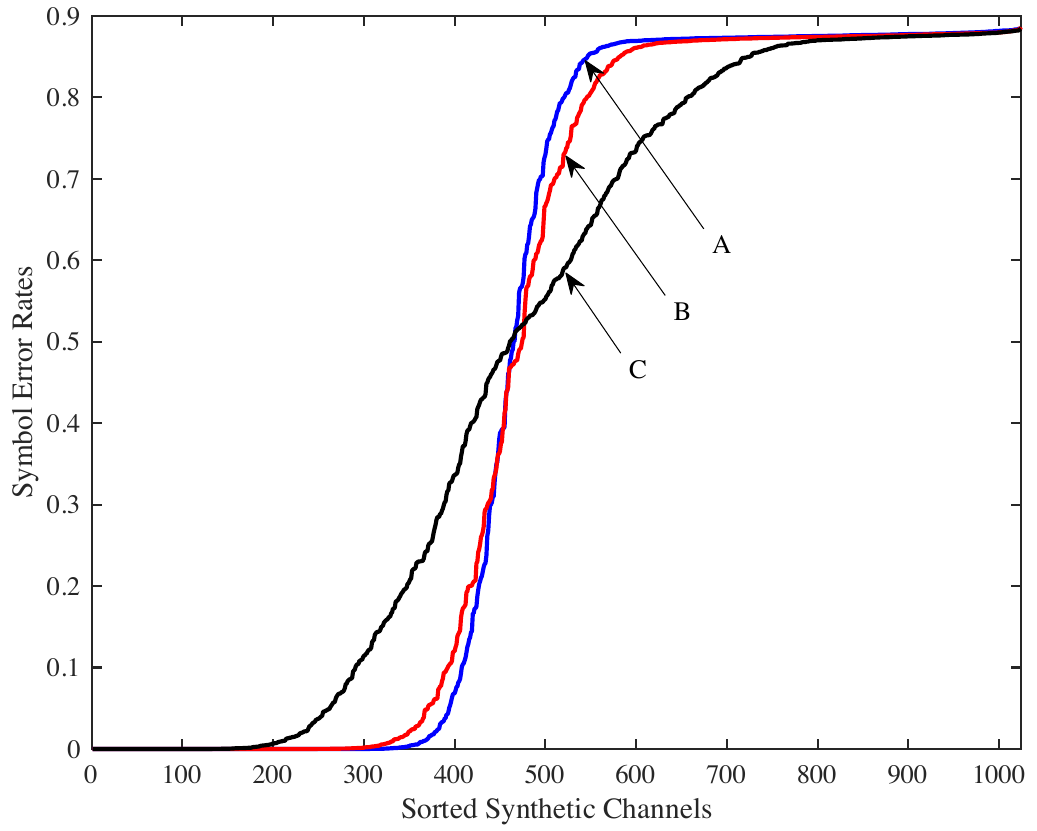}
\caption{The reliabilities of the synthetic channels for $q=8$ non-binary polar codes.} 
\label{Fig_2}
\end{figure} 
\subsection{Review of non-binary polar coding results}
The results of polar coding in \cite{arikan_channel_2009}--\cite{arikan_rate_2009} were generalized to $q$-ary input alphabets in \cite{sasoglu_polarization_2009}--\cite{sasoglu_polar_2011}. 
In one of the approaches proposed in these works, the polarization of the synthetic channels to good or bad channels is guaranteed for channels with input alphabets of prime cardinality $q$ \cite{sasoglu_polarization_2009} in a similar way to \cite{arikan_channel_2009} by using the polar coding kernel with addition mod-$q$ instead of mod-$2$.
To extend these results to channels with input alphabet cardinalities given by composite numbers $q$, a randomized construction was proposed in \cite{sasoglu_polarization_2009} without any change of complexity in the encoder and decoder. The randomization was first used in \cite{arikan_channel_2009} as a process to simplify the analysis of the polarization property.
When $q$ is a composite number, properties of polarizing transforms were investigated in  \cite{sasoglu_polar_2012} and the following result was presented:
\begin{enumerate}[(p.i)]
\item  for any $2\leq K \leq q-1$ and distinct $a_0,\dots,a_{K-1}$, the matrix 
$B_{ij}=f(a_i,a_j),\ i,j=0,\dots,K-1$ of a polarizing mapping has at least $K+1$ distinct entries.
\end{enumerate}

Later, a particular transform $$f(u_1,u_2)=u_1\oplus\pi(u_2),$$ as shown in Fig.~\ref{Fig_1} was provided by \c{S}a\c{s}o\u{g}lu in \cite{sasoglu_polar_2012}--\cite{sasoglu_polarization_2011} with the following definition:  
\begin{equation}
\pi(x)=\left\{
\begin{array}{ll}
\lfloor q/2\rfloor   & \textrm{if } x=0,\\
x-1   & \textrm{if } 1\leq x \leq \lfloor q/2\rfloor,\\
x      & \textrm{otherwise},
\end{array}\right.
\label{polaproc_1}
\end{equation}
as a solution to polarize channels with arbitrary input alphabet sizes. 
However, these types of transforms were not discussed up to now to construct non-binary polar codes with better distance properties. 
In this paper, we provide a particular type of transforms for non-binary polar codes.
In Fig.~\ref{Fig_2}, the reliabilities of the synthetic channels obtained by Monte-Carlo simulation are shown for non-binary polar codes, where A with the proposed transform in this paper, B with transform defined by (\ref{polaproc_1}) and C with standard transform. Here, the block length is $1024$, $q=8$, and  $8$-PSK signalling is used for additive white Gaussian noise (AWGN) channel.
The Fig.~\ref{Fig_2} shows that polarization speed is improved by the use of the introduced transform in this paper.

A similar problem was studied in \cite{abbe_entropies_2015} by Abbe. As an alternative approach, a multi-level code construction technique was also proposed in \cite{sasoglu_polarization_2009} to avoid designing transforms. 
In this regard, $q$-ary multi-level polarization was investigated in \cite{park_polar_2013}--\cite{sahebi_multilevel_2013} independently.
In this case, the synthetic channels converge to good, bad, and \textsl{partially good} channels. 
In \cite{park_multilevel_2011}, it was shown that the sum capacity can be achieved by the multi-level construction. 
Then, it was shown in \cite{park_controlled_2013} that the number of partially polarized levels can be adjusted by using a particular polarizing transform.  

Apart from these research directions, the code construction method proposed in \cite{tal_how_2013}, and which is based on the quantization of the synthetic channels via degrading and upgrading channels, was generalized in \cite{tal_constructing_2012} for arbitrary input alphabets by using successive approximations. 
An alternative code construction method was proposed in \cite{cayci_polar_2012} for arbitrary input alphabets by using one multi dimensional approximation instead of successive approximations. 
The construction for moderate size input alphabets was also discussed in \cite{tal_construction_2015}. 
Then, efficient algorithms were studied in \cite{gulcu_construction_2016} for $q$-ary construction that merge the output symbols of the synthetic channels. 

The rest of the paper is organized as follows. Section \ref{S_system_model} provides a system model of non-binary polar coding for AWGN channel with the encoding scheme and signal sets in 2-dimensions. Furthermore, we propose the minimum distance of the synthetic good channel as a function of the polarizing transform for a given signal set. Section \ref{S_design_polarization} provides a procedure to design polarizing transforms with better distance properties. Moreover, we propose a definition of the equidistant polarizing transforms that achieve the optimal distance spectrum upper bound for a given signal set. Section \ref{S_new_polarization} describes polarizing transforms for $q$-ary PSK signal sets to improve minimum distances and{\slash}or distance spectrums. Moreover, we show that the new polarizing transforms are better than the transforms in (\ref{polaproc_1}) and the standard transform. We propose the equidistant polarizing transform for $q=5$ and $5$-PSK signal set. We reported that the equidistant property is not achievable for $q=4$ by the use of PSK signal set, and hence, Section \ref{S_signalset_design} designs a new signal set for $q=4$ in 2-dimensions in order to define an equidistant polarizing transform.
Section \ref{S_speed_polarization} reviews the polarization speed of the equidistant polarizing transforms.
Section \ref{S_asymptotic} propose asymptotic behaviour of the equidistant property.
Section \ref{S_simulation_result} provides some performance results to investigate the effect of better distance properties by using the equidistant polarizing transforms. Section \ref{S_conclusion} concludes this study with some opinions.

\pagebreak
\section{System Model}\label{S_system_model}
\begin{figure}[hp]
\centering
\begin{tikzpicture}[thick,scale=0.3*1.0, every node/.style={scale=0.45*2*1.0}]
\draw[->]  (-1,0)  -- (-0.2,0) ;
\draw  (0.7,0)  -- (1+1-0.5,0) ;
\draw[->]  (0.25,-2)  -- (0.25,-1.2) ;
\draw  (-1,-2)  -- (1+1-0.5,-2) ;
\draw (-0.2,-1.2)  -- (0.7,-1.2);
\draw (-0.2,0.5)  -- (0.7,0.5);
\draw (-0.2,-1.2)  -- (-0.2,0.5);
\draw (0.7,-1.2)  -- (0.7,0.5);
\begin{scope}[thick]
\draw (0.25,-0.35) node[] {$f$};
\end{scope}
\draw (-1,0) node[left] {$u_1$};
\draw (-1,-2) node[left] {$u_2$};
\draw (1+2-1.5,0)  -- (-1+4.5,0);
\draw (1+2-1.5,-2)  -- (-1+4.5,0-4);
\draw[->]  (-1,0-4)  -- (-0.2,0-4) ;
\draw  (0.7,0-4)  -- (1+1-0.5,0-4) ;
\draw[->]  (0.25,-2-4)  -- (0.25,-1.2-4) ;
\draw  (-1,-2-4)  -- (1+1-0.5,-2-4) ;
\draw (-0.2,-1.2-4)  -- (0.7,-1.2-4);
\draw (-0.2,0.5-4)  -- (0.7,0.5-4);
\draw (-0.2,-1.2-4)  -- (-0.2,0.5-4);
\draw (0.7,-1.2-4)  -- (0.7,0.5-4);
\begin{scope}[thick]
\draw (0.25,-0.35-4) node[] {$f$};
\end{scope}
\draw (-1,0-4) node[left] {$u_3$};
\draw (-1,-2-4) node[left] {$u_4$};
\draw (1+2-1.5,0-4)   -- (-1+4.5,-2);
\draw (1+2-1.5,-2-4)  -- (-1+4.5,-2-4);
\draw[->]  (-1+4.5,0)  -- (-0.2+4.5,0) ;
\draw  (0.7+4.5,0)  -- (1+1-0.5+4.5,0) ;
\draw[->]  (0.25+4.5,-2)  -- (0.25+4.5,-1.2) ;
\draw  (-1+4.5,-2)  -- (1+1-0.5+4.5,-2) ;
\draw (-0.2+4.5,-1.2)  -- (0.7+4.5,-1.2);
\draw (-0.2+4.5,0.5)  -- (0.7+4.5,0.5);
\draw (-0.2+4.5,-1.2)  -- (-0.2+4.5,0.5);
\draw (0.7+4.5,-1.2)  -- (0.7+4.5,0.5);
\begin{scope}[thick]
\draw (0.25+4.5,-0.35) node[] {$f$};
\end{scope}
%
\draw[->]  (-1+4.5,0-4)  -- (-0.2+4.5,0-4) ;
\draw  (0.7+4.5,0-4)  -- (1+1-0.5+4.5,0-4) ;
\draw[->]  (0.25+4.5,-2-4)  -- (0.25+4.5,-1.2-4) ;
\draw  (-1+4.5,-2-4)  -- (1+1-0.5+4.5,-2-4) ;
\draw (-0.2+4.5,-1.2-4)  -- (0.7+4.5,-1.2-4);
\draw (-0.2+4.5,0.5-4)  -- (0.7+4.5,0.5-4);
\draw (-0.2+4.5,-1.2-4)  -- (-0.2+4.5,0.5-4);
\draw (0.7+4.5,-1.2-4)  -- (0.7+4.5,0.5-4);
\begin{scope}[thick]
\draw (0.25+4.5,-0.35-4) node[] {$f$};
\end{scope}
\draw[->]  (-1,0-8)  -- (-0.2,0-8) ;
\draw  (0.7,0-8)  -- (1+1-0.5,0-8) ;
\draw[->]  (0.25,-2-8)  -- (0.25,-1.2-8) ;
\draw  (-1,-2-8)  -- (1+1-0.5,-2-8) ;
\draw (-0.2,-1.2-8)  -- (0.7,-1.2-8);
\draw (-0.2,0.5-8)  -- (0.7,0.5-8);
\draw (-0.2,-1.2-8)  -- (-0.2,0.5-8);
\draw (0.7,-1.2-8)  -- (0.7,0.5-8);
\begin{scope}[thick]
\draw (0.25,-0.35-8) node[] {$f$};
\end{scope}
\draw (-1,0-8) node[left] {$u_5$};
\draw (-1,-2-8) node[left] {$u_6$};
\draw (1+2-1.5,0-8)  -- (-1+4.5,0-8);
\draw (1+2-1.5,-2-8)  -- (-1+4.5,0-4-8);
\draw[->]  (-1,0-4-8)  -- (-0.2,0-4-8) ;
\draw  (0.7,0-4-8)  -- (1+1-0.5,0-4-8) ;
\draw[->]  (0.25,-2-4-8)  -- (0.25,-1.2-4-8) ;
\draw  (-1,-2-4-8)  -- (1+1-0.5,-2-4-8) ;
\draw (-0.2,-1.2-4-8)  -- (0.7,-1.2-4-8);
\draw (-0.2,0.5-4-8)  -- (0.7,0.5-4-8);
\draw (-0.2,-1.2-4-8)  -- (-0.2,0.5-4-8);
\draw (0.7,-1.2-4-8)  -- (0.7,0.5-4-8);
\begin{scope}[thick]
\draw (0.25,-0.35-4-8) node[] {$f$};
\end{scope}
\draw (-1,0-4-8) node[left] {$u_7$};
\draw (-1,-2-4-8) node[left] {$u_8$};
\draw (1+2-1.5,0-4-8)   -- (-1+4.5,-2-8);
\draw (1+2-1.5,-2-4-8)  -- (-1+4.5,-2-4-8);
\draw[->]  (-1+4.5,0-8)  -- (-0.2+4.5,0-8) ;
\draw  (0.7+4.5,0-8)  -- (1+1-0.5+4.5,0-8) ;
\draw[->]  (0.25+4.5,-2-8)  -- (0.25+4.5,-1.2-8) ;
\draw  (-1+4.5,-2-8)  -- (1+1-0.5+4.5,-2-8) ;
\draw (-0.2+4.5,-1.2-8)  -- (0.7+4.5,-1.2-8);
\draw (-0.2+4.5,0.5-8)  -- (0.7+4.5,0.5-8);
\draw (-0.2+4.5,-1.2-8)  -- (-0.2+4.5,0.5-8);
\draw (0.7+4.5,-1.2-8)  -- (0.7+4.5,0.5-8);
\begin{scope}[thick]
\draw (0.25+4.5,-0.35-8) node[] {$f$};
\end{scope}
%
\draw[->]  (-1+4.5,0-4-8)  -- (-0.2+4.5,0-4-8) ;
\draw  (0.7+4.5,0-4-8)  -- (1+1-0.5+4.5,0-4-8) ;
\draw[->]  (0.25+4.5,-2-4-8)  -- (0.25+4.5,-1.2-4-8) ;
\draw  (-1+4.5,-2-4-8)  -- (1+1-0.5+4.5,-2-4-8) ;
\draw (-0.2+4.5,-1.2-4-8)  -- (0.7+4.5,-1.2-4-8);
\draw (-0.2+4.5,0.5-4-8)  -- (0.7+4.5,0.5-4-8);
\draw (-0.2+4.5,-1.2-4-8)  -- (-0.2+4.5,0.5-4-8);
\draw (0.7+4.5,-1.2-4-8)  -- (0.7+4.5,0.5-4-8);
\begin{scope}[thick]
\draw (0.25+4.5,-0.35-4-8) node[] {$f$};
\end{scope}
\draw (1+2-1.5+9/2,0)  -- (-1+4.5+9/2,0);
\draw (1+2-1.5+9/2,-2)  -- (-1+4.5+9/2,0-4);

\draw (1+2-1.5+9/2,0-4)   -- (-1+4.5+9/2,-8);
\draw (1+2-1.5+9/2,-2-4)  -- (-1+4.5+9/2,-2-4-6);
\draw[->]  (-1+4.5+9/2,0)  -- (-0.2+4.5+9/2,0) ;
\draw[->]  (0.7+4.5+9/2,0)  -- (1+1-0.5+4.5+9/2,0) ;
\draw[->]  (0.25+4.5+9/2,-2)  -- (0.25+4.5+9/2,-1.2) ;
\draw[->]  (-1+4.5+9/2,-2)  -- (1+1-0.5+4.5+9/2,-2) ;
\draw (-0.2+4.5+9/2,-1.2)  -- (0.7+4.5+9/2,-1.2);
\draw (-0.2+4.5+9/2,0.5)  -- (0.7+4.5+9/2,0.5);
\draw (-0.2+4.5+9/2,-1.2)  -- (-0.2+4.5+9/2,0.5);
\draw (0.7+4.5+9/2,-1.2)  -- (0.7+4.5+9/2,0.5);
\begin{scope}[thick]
\draw (0.25+4.5+9/2,-0.35) node[] {$f$};
\end{scope}
\draw (1+2-1.5+4.5+9/2,0-0.03) node[right] {$x_1$};
\draw (1+2-1.5+4.5+9/2,-2-0.03) node[right] {$x_2$};
\draw[->]  (-1+4.5+9/2,0-4)  -- (-0.2+4.5+9/2,0-4) ;
\draw[->]  (0.7+4.5+9/2,0-4)  -- (1+1-0.5+4.5+9/2,0-4) ;
\draw[->]  (0.25+4.5+9/2,-2-4)  -- (0.25+4.5+9/2,-1.2-4) ;
\draw[->]  (-1+4.5+9/2,-2-4)  -- (1+1-0.5+4.5+9/2,-2-4) ;
\draw (-0.2+4.5+9/2,-1.2-4)  -- (0.7+4.5+9/2,-1.2-4);
\draw (-0.2+4.5+9/2,0.5-4)  -- (0.7+4.5+9/2,0.5-4);
\draw (-0.2+4.5+9/2,-1.2-4)  -- (-0.2+4.5+9/2,0.5-4);
\draw (0.7+4.5+9/2,-1.2-4)  -- (0.7+4.5+9/2,0.5-4);
\begin{scope}[thick]
\draw (0.25+4.5+9/2,-0.35-4) node[] {$f$};
\end{scope}
\draw (1+2-1.5+4.5+9/2,0-0.03-4) node[right] {$x_3$};
\draw (1+2-1.5+4.5+9/2,-2-0.03-4) node[right] {$x_4$};
\draw (1+2-1.5+9/2,0-8)  -- (-1+4.5+9/2,0-8+6);
\draw (1+2-1.5+9/2,-2-8)  -- (-1+4.5+9/2,0-4-8+6);
\draw (1+2-1.5+9/2,0-4-8)   -- (-1+4.5+9/2,-2-8);
\draw (1+2-1.5+9/2,-2-4-8)  -- (-1+4.5+9/2,-2-4-8);
\draw[->]  (-1+4.5+9/2,0-8)  -- (-0.2+4.5+9/2,0-8) ;
\draw[->]  (0.7+4.5+9/2,0-8)  -- (1+1-0.5+4.5+9/2,0-8) ;
\draw[->]  (0.25+4.5+9/2,-2-8)  -- (0.25+4.5+9/2,-1.2-8) ;
\draw[->]  (-1+4.5+9/2,-2-8)  -- (1+1-0.5+4.5+9/2,-2-8) ;
\draw (-0.2+4.5+9/2,-1.2-8)  -- (0.7+4.5+9/2,-1.2-8);
\draw (-0.2+4.5+9/2,0.5-8)  -- (0.7+4.5+9/2,0.5-8);
\draw (-0.2+4.5+9/2,-1.2-8)  -- (-0.2+4.5+9/2,0.5-8);
\draw (0.7+4.5+9/2,-1.2-8)  -- (0.7+4.5+9/2,0.5-8);
\begin{scope}[thick]
\draw (0.25+4.5+9/2,-0.35-8) node[] {$f$};
\end{scope}
\draw (1+2-1.5+4.5+9/2,0-0.03-8) node[right] {$x_5$};
\draw (1+2-1.5+4.5+9/2,-2-0.03-8) node[right] {$x_6$};
\draw[->]  (-1+4.5+9/2,0-4-8)  -- (-0.2+4.5+9/2,0-4-8) ;
\draw[->]  (0.7+4.5+9/2,0-4-8)  -- (1+1-0.5+4.5+9/2,0-4-8) ;
\draw[->]  (0.25+4.5+9/2,-2-4-8)  -- (0.25+4.5+9/2,-1.2-4-8) ;
\draw[->]  (-1+4.5+9/2,-2-4-8)  -- (1+1-0.5+4.5+9/2,-2-4-8) ;
\draw (-0.2+4.5+9/2,-1.2-4-8)  -- (0.7+4.5+9/2,-1.2-4-8);
\draw (-0.2+4.5+9/2,0.5-4-8)  -- (0.7+4.5+9/2,0.5-4-8);
\draw (-0.2+4.5+9/2,-1.2-4-8)  -- (-0.2+4.5+9/2,0.5-4-8);
\draw (0.7+4.5+9/2,-1.2-4-8)  -- (0.7+4.5+9/2,0.5-4-8);
\begin{scope}[thick]
\draw (0.25+4.5+9/2,-0.35-4-8) node[] {$f$};
\end{scope}
\draw (1+2-1.5+4.5+9/2,0-0.03-4-8) node[right] {$x_7$};
\draw (1+2-1.5+4.5+9/2,-2-0.03-4-8) node[right] {$x_8$};
\end{tikzpicture}
\caption{An encoder scheme of non-binary polar codes.}
\label{Fig_3}
\end{figure}
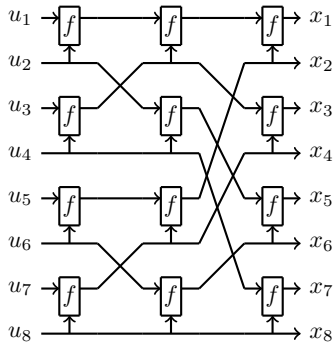

\begin{figure}[hp]
\centering
\begin{tikzpicture}[thick,scale=0.85, every node/.style={scale=0.85}]
\draw  (-1.2-2.75,0)  -- (1.2-2.75,0) ;
\draw  (0-2.75,-1.2)  -- (0-2.75,1.2) ;
\begin{scope}[very thin]
\draw [<->] (0-2.75,0)  -- (0.866-2.75,0.5) ;
\draw  (0.333-2.75,0.25) node[above] {$\begin{smallmatrix}\sqrt{E_s}\end{smallmatrix}$};
\end{scope}
\begin{scope}[very thin,dashed]
\draw (0-2.75,0) circle (1.0cm);
\end{scope}
\draw[fill=gray!160!white] (1-2.75,0) circle (0.05cm) node[above] {$s_0$};
\draw[fill=gray!160!white] (-0.5-2.75,0.866) circle (0.05cm) node[ left] {$s_1$};
\draw[fill=gray!160!white] (-0.5-2.75,-0.866) circle (0.05cm) node[ left] {$s_2$};
\draw  (-1.2,0)  -- (1.2,0) ;
\draw  (0,-1.2)  -- (0,1.2) ;
\begin{scope}[very thin]
\draw[<->]  (0,0)  -- (0.866,0.5) ;
\draw  (0.333,0.25) node[above] {$\begin{smallmatrix}\sqrt{E_s}\end{smallmatrix}$};
\end{scope}
\begin{scope}[very thin,dashed]
\draw (0,0) circle (1.0cm);
\end{scope}
\draw[fill=gray!160!white] (1,0) circle (0.05cm) node[above ] {$s_0$};
\draw[fill=gray!160!white] (0.0,1.0) circle (0.05cm) node[above] {$s_1$};
\draw[fill=gray!160!white] (-1.0,0.0) circle (0.05cm) node[below] {$s_2$};
\draw[fill=gray!160!white] (0.0,-1.0) circle (0.05cm) node[below] {$s_3$};
\draw  (-1.2+2.75,0)  -- (1.2+2.75,0) ;
\draw  (0+2.75,-1.2)  -- (0+2.75,1.2) ;
\begin{scope}[very thin]
\draw[<->]  (0+2.75,0)  -- (0.866+2.75,0.5) ;
\draw  (0.333+2.75,0.25) node[above] {$\begin{smallmatrix}\sqrt{E_s}\end{smallmatrix}$};
\end{scope}
\begin{scope}[very thin,dashed]
\draw (0+2.75,0) circle (1.0cm);
\end{scope}
\draw[fill=gray!160!white] (1+2.75,0) circle (0.05cm) node[above ] {$s_0$};
\draw[fill=gray!160!white] (0.309016994374947 +2.75, 0.951056516295154) circle (0.05cm) node[ right] {$s_1$};
\draw[fill=gray!160!white] (-0.809016994374947 +2.75, 0.587785252292473) circle (0.05cm) node[ left] {$s_2$};
\draw[fill=gray!160!white] (-0.809016994374947+2.75, -0.587785252292473) circle (0.05cm) node[ left] {$s_3$};
\draw[fill=gray!160!white] (0.309016994374947+2.75, -0.951056516295154) circle (0.05cm) node[ right] {$s_4$};
\draw[fill=gray!160!white] (0.0+2.75,-1.20) node[below] {$$};
\end{tikzpicture}
\caption{$q$-ary PSK signal sets: $s_k=\sqrt{E_s}e^{2\pi k/q}\in \mathcal{S}$ for $q=\{3,4,5\}$.}
\label{Fig_4}
\end{figure}
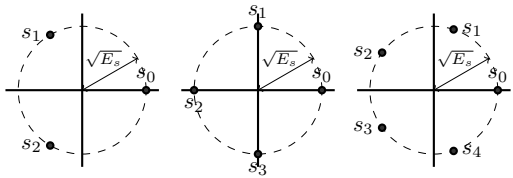

A non-binary encoder scheme with a polarizing transform is depicted in Fig.~\ref{Fig_3} for the block length $N=8$.    
\newline The encoding complexity is the same for any given transform due to the butterfly structure.
In this study, we consider PSK signal sets for $q$-ary input alphabets. 
Here, $u_i \in \mathcal{X}$, where $\mathcal{X}=\{0,1,\dots,q-1\}$. 
Let $\mathcal{S}$ be a signal set with size $q$, where $s_k=\sqrt{E_s}e^{2\pi k/q}\in \mathcal{S}$ for $k=0,1,\dots,q-1$. Here, the signal energy is $E_s$ joule/2-dimensions, and the $q$-ary PSK signal sets $\mathcal{S}:\left\{\sqrt{E_s},\sqrt{E_s} e^{j\frac{2\pi}{q}},\dots,\sqrt{E_s} e^{j\frac{2\pi(q-1)}{q}}\right\}$ are depicted in Fig.~\ref{Fig_4} for different values of $q$.

Here, a natural mapping $x_i\rightarrow s_{x_i}$ is considered for transmission on AWGN channel. 
The noisy observations from the channel are defined for $i=1,\dots,N$ as follows: 
\begin{equation}
{ y_i=s_{x_i}+n_i},
\end{equation} 
where $n_i$ is a complex Gaussian random variable with $\mathcal{CN}(0,\sigma^2)$. 
The power spectral density is $\sigma^2=N_0$ joule/2-dimensions. 
Hence, the signal to noise ratio is ${E_s}/{N_0}$. 

The transition probabilities of the synthetic channels obtained after one-step of polarization are defined as follows:
\begin{eqnarray}
W(y_1,y_2|u_1)=\frac{1}{q}\sum_{u_2=0}^{q-1}W(y_1|f(u_1,u_2))W(y_2|u_2),   \\
W(y_1,y_2,u_1|u_2)=\frac{1}{q}W(y_1|f(u_1,u_2))W(y_2|u_2),
\end{eqnarray}
where $W(y|x)=\frac{1}{\pi \sigma^2}e^{-{\|y-s_x\|^2}/{\sigma^2}}$. These transition probabilities describe a link between $f$ and error performances for polarized channels.  Note that $\sigma^2$ is the variance for 2-dimensions. Squared \textsl{Euclidean} distance is denoted by $\|\cdot\|^2$.

For the bad channel, the distance is $d=\|s_{k}-s_{k+1}\|$. For the synthetic good channel, the distance of a given transform $f$ is 
\begin{eqnarray}
d=\sqrt{\|s_{f(u_1,u_2)}-s_{f(u_1,u_2')}\|^2+\|s_{u_2}-s_{u_2'}\|^2}
\end{eqnarray} 
for any $u_1$ and $u_2\neq u_2'$. The minimum distance of the standard transform is given as follows:
\begin{eqnarray}
d_{min}=\sqrt{\|s_{m}-s_{m+1}\|^2+\|s_{n}-s_{n+1}\|^2},
\label{std_dmin}
\end{eqnarray} 
for any $m$, $n$. The standard transform provides the minimum distance for PSK signal set as the following equation.
\begin{eqnarray}
d_{min}&=&2\sqrt{2}\sin\left({\pi}/{q}\right)\sqrt{E_s}
\label{mindist}
\end{eqnarray}

There are some results that we can consider here as follows:
\begin{itemize}
\item Standard transform provides $d_{min}$ in (\ref{std_dmin}) which is the {\bf minimum} $d_{min}$ that can be obtained for a given transform.
\item Transform in (\ref{polaproc_1}) provides that there is one signal which is more protected than others and there are $q-1$ signals with the same $d_{min}$ in (\ref{std_dmin}). Hence, effective minimum distance is bigger than (\ref{std_dmin}). An example is provided in the section for polarizing transforms for $q=8$. As a result, the asymptotic behaviour of transform in (\ref{polaproc_1}) is close to the standard transform. For this case, we will discuss the simulation results in Fig.\ref{fig_fer1} and Fig.\ref{fig_fer3}.  
\end{itemize}

In the next section, we will describe equidistant transforms which improve distance properties.

\section{Design of Polarizing Transforms}\label{S_design_polarization}
For the AWGN channel, the distance spectrum upper bound is given for the probability of symbol error $P_e$ for a codebook with distance spectrum $N(d)$ as follows:
\begin{eqnarray}
P_e\leq\sum_{d\geq d_{min}} N(d)\cdot Q\left({d}/{(2\sigma)}\right),
\label{dsb}
\end{eqnarray}  
where $Q(d/(2\sigma))$ corresponds to the pairwise error probability between two points at a distance $d$ apart, and $N(d)$ denotes the distance spectrum defined as the number of points at distance $d$\footnote{Obviously, $Q(x)=\frac{1}{\sqrt{2\pi}}\int_{x}^{\infty} e^{-t^2/2} dt$  for the channel with $\sigma=1$. Notice that $\sigma^2$ is for 1-dimension in (\ref{dsb}), and $SNR=E_s/N_0$ where the signal power is $E_s$ joule/2-dimensions and $N_0$ is $\sigma^2$ joule/2-dimensions.}. A book-context for distance spectrum bound can be found in \cite{RamZamir}.

We investigate the distance properties by using a table where $u_1$ and $u_2$ are shown in a cell with coordinates $(x_1,x_2)$  corresponding to the outputs of the scheme in Fig.~\ref{Fig_1}. For more clarity, the cells are marked with a grey face for $u_1=0$.
We give a table for the standard transform (i.e. with the type of $f(u_1,u_2)=u_1\oplus \pi_0(u_2)$, where $\pi_0$ is the identical permutation) for $q=5$ as follows:
$$
\begin{tikzpicture}[scale=0.85, every node/.style={scale=0.85}]
\draw  (-1.5,0)  -- (1.5,0) ;
\draw  (-1.0,-0.5)  -- (1.5,-0.5) ;
\draw  (-1.0,-1.0)  -- (1.5,-1.0) ;
\draw  (-1.0,-1.5)  -- (1.5,-1.5) ;
\draw  (-1.0,-2.0)  -- (1.5,-2.0) ;
\draw  (-1.0,-2.5)  -- (1.5,-2.5) ;
\draw  (-1.0, 0.5)  -- (-1.0,-2.5) ;
\draw  (-0.5, 0.0)  -- (-0.5,-2.5) ;
\draw  (-0.0, 0.0)  -- (-0.0,-2.5) ;
\draw  (0.5, 0.0)  -- (0.5,-2.5) ;
\draw  (1.0, 0.0)  -- (1.0,-2.5) ;
\draw  (1.5, 0.0)  -- (1.5,-2.5) ;
\draw  (-1.5, 0.5)  -- (-1.0,0.0) ;
\draw (-1.2,0.5) node[] {$x_1$};
\draw (-0.75,0) node[above] {$0$};
\draw (-0.25,0) node[above] {$1$};
\draw ( 0.25,0) node[above] {$2$};
\draw ( 0.75,0) node[above] {$3$};
\draw ( 1.25,0) node[above] {$4$};
\draw (-1.5,-0.0) node[above] {$x_2$};
\draw (-1.25,-0.5) node[above] {$0$};
\draw (-1.25,-1.0) node[above] {$1$};
\draw (-1.25,-1.5) node[above] {$2$};
\draw (-1.25,-2.0) node[above] {$3$};
\draw (-1.25,-2.5) node[above] {$4$};
\draw[fill=gray!50] (-1.25+0.5-0.25,-0.5) rectangle (-1.25+0.5+0.25,-0.5+0.5) (-1.25+0.5,-0.5) node[above] {$00$};
\draw (-1.25+0.5,-1.0) node[above] {$41$};
\draw (-1.25+0.5,-1.5) node[above] {$32$};
\draw (-1.25+0.5,-2.0) node[above] {$23$};
\draw (-1.25+0.5,-2.5) node[above] {$14$};
\draw (-1.25+1.0,-0.5) node[above] {$10$};
\draw[fill=gray!50] (-1.25+1.0-0.25,-1.0) rectangle (-1.25+1.0+0.25,-1.0+0.5) (-1.25+1.0,-1.0) node[above] {$01$};
\draw (-1.25+1.0,-1.5) node[above] {$42$};
\draw (-1.25+1.0,-2.0) node[above] {$33$};
\draw (-1.25+1.0,-2.5) node[above] {$24$};
\draw (-1.25+1.5,-0.5) node[above] {$20$};
\draw (-1.25+1.5,-1.0) node[above] {$11$};
\draw[fill=gray!50] (-1.25+1.5-0.25,-1.5) rectangle (-1.25+1.5+0.25,-1.5+0.5) (-1.25+1.5,-1.5) node[above] {$02$};
\draw (-1.25+1.5,-2.0) node[above] {$43$};
\draw (-1.25+1.5,-2.5) node[above] {$34$};
\draw (-1.25+2.0,-0.5) node[above] {$30$};
\draw (-1.25+2.0,-1.0) node[above] {$21$};
\draw (-1.25+2.0,-1.5) node[above] {$12$};
\draw[fill=gray!50] (-1.25+2.0-0.25,-2.0) rectangle (-1.25+2.0+0.25,-2.0+0.5) (-1.25+2.0,-2.0) node[above] {$03$};
\draw (-1.25+2.0,-2.5) node[above] {$44$};
\draw (-1.25+2.5,-0.5) node[above] {$40$};
\draw (-1.25+2.5,-1.0) node[above] {$31$};
\draw (-1.25+2.5,-1.5) node[above] {$22$};
\draw (-1.25+2.5,-2.0) node[above] {$13$};
\draw[fill=gray!50] (-1.25+2.5-0.25,-2.5) rectangle (-1.25+2.5+0.25,-2.5+0.5) (-1.25+2.5,-2.5) node[above] {$04$};
\draw (0.25,-2.5) node[below] {$\pi_0=\left(\begin{smallmatrix} 0& 1& 2& 3& 4\\ 0&1&2&3&4 \end{smallmatrix}\right).$};
\end{tikzpicture}$$

\begin{myexample}
Suppose that $(u_1=0,u_2=2)$ and hence $(s_{x_1=2},s_{x_2=2})$ are transmitted. 
For the decision of $u_2$, the following error events can occur:
\begin{enumerate}
\item There are two error events within squared distance $d^2=2\|s_0-s_1\|^2$ that $(x_1,x_2)\in\{(1,1),(3,3)\}$.
\item There are two more error events within the distance $d^2=2\|s_0-s_2\|^2$ that $(x_1,x_2)\in\{(0,0),(4,4)\}$.
\end{enumerate}
\end{myexample}

The minimum distance of the standard transform is 
\begin{equation}
d_{\min}=\sqrt{2}\|s_0-s_1\|=1.66\sqrt{E_s}
\label{dmin_st}
\end{equation} 
for 5-PSK signalling.
The distance spectrum $N(d)$ for the standard transform is given as follows:   
\begin{equation}
N(d)=\left\{
\begin{array}{ll}
2   & d=1.66\sqrt{E_s},\\
2   & d=\sqrt{2}\|s_0-s_2\|=2.69\sqrt{E_s},\\
0   & otherwise.
\end{array}\right.
\end{equation} 
An analytical expression of the distance spectrum upper bound for the synthetic good channel with the standard transform is  
\begin{eqnarray}
&P_{e}\leq2Q\left(1.66\sqrt{{SNR}/{2}}\right)+2Q\left(2.69\sqrt{{SNR}/{2}}\right).&
\label{eqn_ub}
\end{eqnarray} 
The standard transform is the worst case for distance spectrum upper bound.

As we can see from Fig.~\ref{fig_4}, the analytic result (\ref{eqn_ub}) is a tight upper bound for the symbol error probability. 

\pagebreak
We investigate the distance properties of the transform in (\ref{polaproc_1}) for $q=8$ as follows:
\begin{myexample}
We give the following table for the transform in (\ref{polaproc_1}) with $q=8$.
\begin{center}
\begin{tikzpicture}[scale=0.95, every node/.style={scale=0.95}]
\draw  (-1.5, 0.0)  -- (3, 0.0) ;
\draw  (-1.0,-0.5)  -- (3,-0.5) ;
\draw  (-1.0,-1.0)  -- (3,-1.0) ;
\draw  (-1.0,-1.5)  -- (3,-1.5) ;
\draw  (-1.0,-2.0)  -- (3,-2.0) ;
\draw  (-1.0,-2.5)  -- (3,-2.5) ;
\draw  (-1.0,-3.0)  -- (3,-3.0) ;
\draw  (-1.0,-3.5)  -- (3,-3.5) ;
\draw  (-1.0,-4.0)  -- (3,-4.0) ;
\draw  (-1.0, 0.5)  -- (-1.0,-4) ;
\draw  (-0.5, 0.0)  -- (-0.5,-4) ;
\draw  (-0.0, 0.0)  -- (-0.0,-4) ;
\draw  ( 0.5, 0.0)  -- ( 0.5,-4) ;
\draw  ( 1.0, 0.0)  -- ( 1.0,-4) ;
\draw  ( 1.5, 0.0)  -- ( 1.5,-4) ;
\draw  ( 2.0, 0.0)  -- ( 2.0,-4) ;
\draw  ( 2.5, 0.0)  -- ( 2.5,-4) ;
\draw  ( 3.0, 0.0)  -- ( 3.0,-4) ;
\draw  (-1.5, 0.5)  -- (-1.0,0.0) ;
\draw (-1.2,0.5) node[] {$x_1$};
\draw (-0.75,0) node[above] {$0$};
\draw (-0.25,0) node[above] {$1$};
\draw ( 0.25,0) node[above] {$2$};
\draw ( 0.75,0) node[above] {$3$};
\draw ( 1.25,0) node[above] {$4$};
\draw ( 1.75,0) node[above] {$5$};
\draw ( 2.25,0) node[above] {$6$};
\draw ( 2.75,0) node[above] {$7$};
\draw (-1.5,-0.0) node[above] {$x_2$};
\draw (-1.25,-0.5) node[above] {$0$};
\draw (-1.25,-1.0) node[above] {$1$};
\draw (-1.25,-1.5) node[above] {$2$};
\draw (-1.25,-2.0) node[above] {$3$};
\draw (-1.25,-2.5) node[above] {$4$};
\draw (-1.25,-3.0) node[above] {$5$};
\draw (-1.25,-3.5) node[above] {$6$};
\draw (-1.25,-4.0) node[above] {$7$};
\draw[fill=gray!0] (-1.25+0.5-0.25,-0.5) rectangle (-1.25+0.5-0.25+0.5,-0.5+0.5) (-1.25+0.5+0.25-0.25,-0.5) node[above]  {$40$};%
\draw[fill=gray!50] (-1.25+0.5-0.25,-1.0) rectangle (-1.25+0.5-0.25+0.5,-1.0+0.5) (-1.25+0.5,-1.0) node[above] {$01$};
\draw (-1.25+0.5,-1.5) node[above] {$72$};
\draw (-1.25+0.5,-2.0) node[above] {$63$};
\draw (-1.25+0.5,-2.5) node[above] {$54$};
\draw (-1.25+0.5,-3.0) node[above] {$35$};
\draw (-1.25+0.5,-3.5) node[above] {$26$};
\draw (-1.25+0.5,-4.0) node[above] {$17$};%
\draw (-1.25+1.0,-0.5) node[above] {$50$};
\draw (-1.25+1.0,-1.0) node[above] {$11$};
\draw[fill=gray!50] (-1.25+1.0-0.25,-1.5) rectangle (-1.25+1.0-0.25+0.5,-1.5+0.5) (-1.25+1.0+0.25-0.25,-1.5) node[above]  {$02$};
\draw[fill=gray!0] (-1.25+1.0-0.25,-2.0) rectangle (-1.25+1.0-0.25+0.5,-2.0+0.5) (-1.25+1.0+0.25-0.25,-2.0) node[above]  {$73$};%
\draw (-1.25+1.0,-2.5) node[above]  {$64$};
\draw (-1.25+1.0,-3.0) node[above]  {$45$};
\draw (-1.25+1.0,-3.5) node[above]  {$36$};
\draw (-1.25+1.0,-4.0) node[above]  {$27$};%
\draw (-1.25+1.5,-0.5) node[above] {$60$};
\draw (-1.25+1.5,-1.0) node[above]  {$21$};
\draw (-1.25+1.5,-1.5) node[above] {$12$};
\draw[fill=gray!50] (-1.25+1.5-0.25,-2.0) rectangle (-1.25+1.5-0.25+0.5,-2+0.5) (-1.25+1.5+0.25-0.25,-2) node[above] {$03$};
\draw (-1.25+1.5,-2.5) node[above] {$74$};
\draw (-1.25+1.5,-3.0) node[above] {$55$};
\draw[fill=gray!0] (-1.25+1.5-0.25,-3.5) rectangle (-1.25+1.5-0.25+0.5,-3.5+0.5) (-1.25+1.5+0.25-0.25,-3.5) node[above] {$46$};%
\draw (-1.25+1.5,-4.0) node[above] {$37$};%
\draw (-1.25+2,-0.5) node[above]    {$70$};
\draw[fill=gray!0] (-1.25+2-0.25,-1.0) rectangle (-1.25+2-0.25+0.5,-1.0+0.5) (-1.25+2+0.25-0.25,-1.0) node[above]     {$31$};%
\draw (-1.25+2,-1.5) node[above]    {$22$};
\draw (-1.25+2,-2.0) node[above]    {$13$};
\draw[fill=gray!50] (-1.25+2-0.25,-2.5) rectangle (-1.25+2-0.25+0.5,-2.5+0.5) (-1.25+2+0.25-0.25,-2.5) node[above]    {$04$};
\draw (-1.25+2,-3.0) node[above]    {$65$};
\draw (-1.25+2,-3.5) node[above]    {$56$};
\draw (-1.25+2,-4.0) node[above]    {$47$};%
\draw[fill=gray!50]  (-1.25+2.5-0.25,-0.5) rectangle (-1.25+2.5-0.25+0.5,-0.5+0.5) (-1.25+2.5+0.25-0.25,-0.5) node[above] {$00$};
\draw (-1.25+2.5,-1.0) node[above]  {$41$};
\draw (-1.25+2.5,-1.5) node[above] {$32$};
\draw (-1.25+2.5,-2.0) node[above] {$23$};
\draw[fill=gray!0] (-1.25+2.5-0.25,-2.5) rectangle (-1.25+2.5-0.25+0.5,-2.5+0.5) (-1.25+2.5+0.25-0.25,-2.5) node[above] {$14$};%
\draw (-1.25+2.5,-3.0) node[above] {$75$};
\draw (-1.25+2.5,-3.5) node[above] {$66$};
\draw (-1.25+2.5,-4.0) node[above] {$57$};%
\draw (-1.25+3,-0.5) node[above]    {$10$};
\draw (-1.25+3,-1.0) node[above]     {$51$};
\draw (-1.25+3,-1.5) node[above]    {$42$};
\draw (-1.25+3,-2.0) node[above]    {$33$};
\draw (-1.25+3,-2.5) node[above]    {$24$};
\draw[fill=gray!50] (-1.25+3-0.25,-3.0) rectangle (-1.25+3-0.25+0.5,-3.0+0.5) (-1.25+3+0.25-0.25,-3.0) node[above]    {$05$};
\draw (-1.25+3,-3.5) node[above]    {$76$};
\draw[fill=gray!0] (-1.25+3-0.25,-4.0) rectangle (-1.25+3-0.25+0.5,-4.0+0.5) (-1.25+3+0.25-0.25,-4.0) node[above]    {$67$};%
\draw (-1.25+3.5,-0.5) node[above] {$20$};
\draw (-1.25+3.5,-1.0) node[above]  {$61$};
\draw[fill=gray!0] (-1.25+3.5-0.25,-1.5) rectangle (-1.25+3.5-0.25+0.5,-1.5+0.5) (-1.25+3.5+0.25-0.25,-1.5) node[above] {$52$};%
\draw (-1.25+3.5,-2.0) node[above] {$43$};
\draw (-1.25+3.5,-2.5) node[above] {$34$};
\draw (-1.25+3.5,-3.0) node[above] {$15$};
\draw[fill=gray!50]  (-1.25+3.5-0.25,-3.5) rectangle (-1.25+3.5-0.25+0.5,-3.5+0.5) (-1.25+3.5+0.25-0.25,-3.5) node[above] {$06$};
\draw (-1.25+3.5,-4.0) node[above] {$77$};%
\draw (-1.25+4,-0.5) node[above]    {$30$};
\draw (-1.25+4,-1.0) node[above]     {$71$};
\draw (-1.25+4,-1.5) node[above]    {$62$};
\draw (-1.25+4,-2.0) node[above]    {$53$};
\draw (-1.25+4,-2.5) node[above]    {$44$};
\draw[fill=gray!0] (-1.25+4-0.25,-3.0) rectangle (-1.25+4-0.25+0.5,-3.0+0.5) (-1.25+4+0.25-0.25,-3.0) node[above]    {$25$};%
\draw (-1.25+4,-3.5) node[above]    {$16$};
\draw[fill=gray!50]  (-1.25+4-0.25,-4.0) rectangle (-1.25+4-0.25+0.5,-4.0+0.5) (-1.25+4+0.25-0.25,-4.0) node[above]    {$07$};
\draw(1.0,-4.50) node[] {$\pi=\left(\begin{smallmatrix} 0& 1& 2& 3& 4& 5& 6& 7\\ 4&0&1&2&3&5&6&7 \end{smallmatrix}\right)$};
\end{tikzpicture}
\end{center}
Suppose that $u_1=0$ and $({x_1},{x_2})\in\{(0,1),(1,2),(2,3),(3,4),(4,0),(5,5),(6,6),(7,7)\}$ are encoded to transmit.
For $u_2$ decision, following error events can occur:
\begin{enumerate}
\item There is one error for $u_2=1$ within $d=1.082\sqrt{E_s}$ by $(x_1,x_2)\in\{(1,2)\}$.
\item There are two errors for $u_2=2$ within $d=1.082\sqrt{E_s}$ by $(x_1,x_2)\in\{(0,1),(2,3)\}$. 
\item There are two errors for $u_2=3$ within $d=1.082\sqrt{E_s}$ by $(x_1,x_2)\in\{(1,2),(3,4)\}$.
\item There is one errors for $u_2=4$ within $d=1.082\sqrt{E_s}$ by $(x_1,x_2)\in\{(2,3)\}$.
\item There are three errors for $u_2=0$ within $d=2\sqrt{E_s}$ by $(x_1,x_2)\in\{(5,5),(6,6),(7,7)\}$.
\item There is one error for $u_2=5$ within $d=1.082\sqrt{E_s}$ by $(x_1,x_2)\in\{(6,6)\}$.
\item There are two errors for $u_2=6$ within $d=1.082\sqrt{E_s}$ by $(x_1,x_2)\in\{(5,5),(7,7)\}$.
\item There is one error for $u_2=7$ within $d=1.082\sqrt{E_s}$ by $(x_1,x_2)\in\{(6,6)\}$.
\end{enumerate}
Notice that $u_2=0$ is more protected than others. Then, the effective minimum distance is $d_{min}^{eff}=\frac{7\times1.082+2}{8}\sqrt{E_s}=1.197\sqrt{E_s}$. 
\end{myexample}

In this paper, we first focus on designing polarizing transforms which increase the minimum distance for $q$-ary PSK signal sets. 
To design a table of transform with the type of $f(u_1,u_2)=u_1\oplus \pi(u_2)$, we define a simple procedure as follows: 
\begin{enumerate}[(p.i)]
\item Each row has only 1 candidate of $u_2$ for $u_1=0$,  
\item Each column has only 1 candidate of $u_2$ for $u_1=0$,
\item Place all candidates of $u_2$ as far as from each others for $u_1=0$,
\item Fill the empty cells by $q$ candidates of $u_2$ that are placed in the $k$th cyclic right-shift cell for $u_1=k$, where $k=1,\dots,q-1$.   
\end{enumerate} 
Hence, the completed table provides the polarizing transform with the type of $f(u_1,u_2)=u_1\oplus \pi(u_2)$.
Here, (p.i) and (p.ii) are constraints of the procedure to guarantee $(x_1,x_2)$ that can take all possible $q$-ary pairs. 
To achieve an increased minimum distance, (p.iii) can be done by a computer search. Then, (p.iv) is to complete the definition of the polarizing transform.

For a given finite $q$, polarization transforms with type $f(u_1,u_2)=u_1\oplus \pi(u_2)$ can be written by $q!$ possible permutations. There are also many {\bf equivalent} transforms with type $f(u_1,u_2)=u_1\oplus \pi(u_2)$ for a given signal set. To reduce the number of transforms in the search process, permutations with $\pi(0)=0$ can be considered. Then, a smart search strategy can be used for $(q-1)!$ transforms. We note that the standard transform is the worst and the equidistant transform is the optimal. The main task in the search process is to make the best possible distance profile for a given signal set. Start from the initial case with the candidates far from each other could be a good strategy. Because, when we found an equidistant transform, we can terminate the search earlier. Additionally, an efficient tree search can be applied on the search process of transforms such as sphere decoding that the initial trial on search tree is a factor on the complexity. We consider the asymptotic behaviour of the equidistant property for high order input alphabet sizes in Section \ref{S_asymptotic}. 

\begin{mylemma}
The distance is conserved as follows:
$$\sum_{u_2'}{(\|s_{f(u_1,u_2)}-s_{f(u_1,u_2')}\|^2+\|s_{u_2}-s_{u_2'}\|^2)}
=2\sum_{k=1}^{q-1}\|s_k-s_0\|^2$$
for $q$-ary PSK signal set by using a polarizing transform $f$. {\em Lemma~1} provides that the sum of distances is independent of the polarizing transform $f$ and it is constant.
\end{mylemma}
\begin{myproof}
It is obtained by the following steps:
\begin{itemize}
\item PSK is a signal set that is matched to a group \cite{loeliger_signal_1991} that $\|s_{l+k}-s_l\|=\|s_k-s_0\|$.
\item $\sum_{u_2'}{\|s_{u_2}-s_{u_2'}\|^2}=\sum_{k=1}^{q-1}\|s_k-s_0\|^2.$
\item For a fixed $u_1$, $u_2 \rightarrow f(u_1,u_2)$ is an invertible function of $u_2$. Hence, $\sum_{u_2'}{\|s_{f(u_1,u_2)}-s_{f(u_1,u_2')}\|^2}=\sum_{k=1}^{q-1}\|s_k-s_0\|^2.$
\end{itemize}
\end{myproof}

\begin{mylemma}
The minimum distance of a $q$-ary synthetic good channel is lower bounded by $$d_{min}\leq\sqrt{\frac{2}{q-1}\sum_{k=1}^{q-1}\|s_k-s_0\|^2},$$ where $s_k \in S$, and $S$ is the $q$-ary PSK signal set.
\end{mylemma}
\begin{myproof}
The distance is conserved which is seen in the previous result, and it is easy to see that 
$$\min \{d^2\}\leq \mathcal{D}/(q-1),$$
where $$\mathcal{D}=2\sum_{k=1}^{q-1}\|s_k-s_0\|^2.$$ By this way, the proof of $d_{min}\leq\sqrt{\mathcal{D}/(q-1)}$ is obtained.
\end{myproof}

\begin{myproposition}\label{theo3}
For $a> 0$, $b>0$ and $a\neq b$,
$$2Q\left(\sqrt{\frac{a^2+b^2}{2}}\right) < Q\left(a\right)+Q\left(b\right).$$ 
\label{ww}
\end{myproposition}
The proof of the proposition is provided in Appendix.

\begin{mydefinition}[Equidistant Polarizing Transforms]
For a given $q$-ary signal set the polarizing transforms with the distance spectrum $N(d_{min})=q-1$ are the Equidistant polarizing transforms.
\end{mydefinition}

The equidistant channels was first defined by Jelinek in \cite{jelinek_evaluation_1968}.

\begin{mytheorem}
An equidistant transform for $q$-ary signal set has the distance spectrum upper bound as follows:
$$P_e\leq (q-1)Q(d_{min}/(2\sigma)),$$
where the minimum distance $$d_{min}=\sqrt{\frac{2}{q-1}\sum_{k=1}^{q-1}\|s_k-s_0\|^2}.$$ This is the best achievable distance spectrum bound. 
\label{www}
\end{mytheorem}

\begin{myproof}
The proof is obtained by Lemma 1, Lemma 2 and Proposition 1.
\end{myproof}

In the next section, we provide some examples for the better distance characteristics.

\section{Polarizing Transforms}\label{S_new_polarization}

\begin{myexample}[Equidistant Polarizing Transforms for $q=5$]
We applied the procedure for 5-PSK signal set to design polarizing transforms with the type of $f(u_1,u_2)=u_1\oplus \pi(u_2)$ for $q=5$. 
Hence, the following tables are provided for $q=5$ and 5-PSK signal set. 

$$
\begin{tikzpicture}[scale=0.85, every node/.style={scale=0.85}]
\draw  (-1.5,0)  -- (1.5,0) ;
\draw  (-1.0,-0.5)  -- (1.5,-0.5) ;
\draw  (-1.0,-1.0)  -- (1.5,-1.0) ;
\draw  (-1.0,-1.5)  -- (1.5,-1.5) ;
\draw  (-1.0,-2.0)  -- (1.5,-2.0) ;
\draw  (-1.0,-2.5)  -- (1.5,-2.5) ;
\draw  (-1.0, 0.5)  -- (-1.0,-2.5) ;
\draw  (-0.5, 0.0)  -- (-0.5,-2.5) ;
\draw  (-0.0, 0.0)  -- (-0.0,-2.5) ;
\draw  (0.5, 0.0)  -- (0.5,-2.5) ;
\draw  (1.0, 0.0)  -- (1.0,-2.5) ;
\draw  (1.5, 0.0)  -- (1.5,-2.5) ;
\draw  (-1.5, 0.5)  -- (-1.0,0.0) ;
\draw (-1.2,0.5) node[] {$x_1$};
\draw (-0.75,0) node[above] {$0$};
\draw (-0.25,0) node[above] {$1$};
\draw ( 0.25,0) node[above] {$2$};
\draw ( 0.75,0) node[above] {$3$};
\draw ( 1.25,0) node[above] {$4$};
\draw (-1.5,-0.0) node[above] {$x_2$};
\draw (-1.25,-0.5) node[above] {$0$};
\draw (-1.25,-1.0) node[above] {$1$};
\draw (-1.25,-1.5) node[above] {$2$};
\draw (-1.25,-2.0) node[above] {$3$};
\draw (-1.25,-2.5) node[above] {$4$};
\draw[fill=gray!50] (-1.25+0.5-0.25,-0.5) rectangle (-1.25+0.5+0.25,-0.5+0.5) (-1.25+0.5,-0.5) node[above] {$00$};
\draw (-1.25+0.5,-1.0) node[above] {$31$};
\draw (-1.25+0.5,-1.5) node[above] {$12$};
\draw (-1.25+0.5,-2.0) node[above] {$43$};
\draw (-1.25+0.5,-2.5) node[above] {$24$};
\draw (-1.25+1.0,-0.5) node[above] {$10$};
\draw (-1.25+1.0,-1.0) node[above] {$41$};
\draw (-1.25+1.0,-1.5) node[above] {$22$};
\draw[fill=gray!50] (-1+0.5,-0.5-1.5) rectangle (-0.5+0.5,0-1.5) (-1.25+1.0,-2.0) node[above] {$03$};
\draw (-1.25+1.0,-2.5) node[above] {$34$};
\draw (-1.25+1.5,-0.5) node[above] {$20$};
\draw[fill=gray!50] (-1+1,-0.5-0.5) rectangle (-0.5+1,0-0.5) (-1.25+1.5,-1.0) node[above] {$01$};
\draw (-1.25+1.5,-1.5) node[above] {$32$};
\draw (-1.25+1.5,-2.0) node[above] {$13$};
\draw (-1.25+1.5,-2.5) node[above] {$44$};
\draw (-1.25+2.0,-0.5) node[above] {$30$};
\draw (-1.25+2.0,-1.0) node[above] {$11$};
\draw (-1.25+2.0,-1.5) node[above] {$42$};
\draw (-1.25+2.0,-2.0) node[above] {$23$};
\draw[fill=gray!50] (-1+1.5,-0.5-2.0) rectangle (-0.5+1.5,0-2.0)  (-1.25+2.0,-2.5) node[above] {$04$};
\draw (-1.25+2.5,-0.5) node[above] {$40$};
\draw (-1.25+2.5,-1.0) node[above] {$21$};
\draw[fill=gray!50] (-1+2,-0.5-1.0) rectangle (-0.5+2,0-1.0) (-1.25+2.5,-1.5) node[above] {$02$};
\draw (-1.25+2.5,-2.0) node[above] {$33$};
\draw (-1.25+2.5,-2.5) node[above] {$14$};
\draw (0.25,-2.5) node[below] {$\pi_1=\left(\begin{smallmatrix} 0& 1& 2& 3& 4\\ 0&2&4&1&3 \end{smallmatrix}\right)$};
\end{tikzpicture}
\begin{tikzpicture}[scale=0.85, every node/.style={scale=0.85}]
\draw  (-1.5,0)  -- (1.5,0) ;
\draw  (-1.0,-0.5)  -- (1.5,-0.5) ;
\draw  (-1.0,-1.0)  -- (1.5,-1.0) ;
\draw  (-1.0,-1.5)  -- (1.5,-1.5) ;
\draw  (-1.0,-2.0)  -- (1.5,-2.0) ;
\draw  (-1.0,-2.5)  -- (1.5,-2.5) ;
\draw  (-1.0, 0.5)  -- (-1.0,-2.5) ;
\draw  (-0.5, 0.0)  -- (-0.5,-2.5) ;
\draw  (-0.0, 0.0)  -- (-0.0,-2.5) ;
\draw  (0.5, 0.0)  -- (0.5,-2.5) ;
\draw  (1.0, 0.0)  -- (1.0,-2.5) ;
\draw  (1.5, 0.0)  -- (1.5,-2.5) ;
\draw  (-1.5, 0.5)  -- (-1.0,0.0) ;
\draw (-1.2,0.5) node[] {$x_1$};
\draw (-0.75,0) node[above] {$0$};
\draw (-0.25,0) node[above] {$1$};
\draw ( 0.25,0) node[above] {$2$};
\draw ( 0.75,0) node[above] {$3$};
\draw ( 1.25,0) node[above] {$4$};
\draw (-1.5,-0.0) node[above] {$x_2$};
\draw (-1.25,-0.5) node[above] {$0$};
\draw (-1.25,-1.0) node[above] {$1$};
\draw (-1.25,-1.5) node[above] {$2$};
\draw (-1.25,-2.0) node[above] {$3$};
\draw (-1.25,-2.5) node[above] {$4$};
\draw[fill=gray!50] (-1.25+0.5-0.25,-0.5) rectangle (-1.25+0.5+0.25,-0.5+0.5) (-1.25+0.5,-0.5) node[above] {$00$};
\draw (-1.25+0.5,-1.0) node[above] {$21$};
\draw (-1.25+0.5,-1.5) node[above] {$42$};
\draw (-1.25+0.5,-2.0) node[above] {$13$};
\draw (-1.25+0.5,-2.5) node[above] {$34$};
\draw (-1.25+1.0,-0.5) node[above] {$10$};
\draw (-1.25+1.0,-1.0) node[above] {$31$};
\draw[fill=gray!50] (-1.25+1.0-0.25,-1.5) rectangle (-1.25+1.0+0.25,-1.5+0.5) (-1.25+1.0,-1.5) node[above] {$02$};
\draw (-1.25+1.0,-2.0) node[above] {$23$};
\draw (-1.25+1.0,-2.5) node[above] {$44$};
\draw (-1.25+1.5,-0.5) node[above] {$20$};
\draw (-1.25+1.5,-1.0) node[above] {$41$};
\draw (-1.25+1.5,-1.5) node[above] {$12$};
\draw (-1.25+1.5,-2.0) node[above] {$33$};
\draw[fill=gray!50] (-1.25+1.5-0.25,-2.5) rectangle (-1.25+1.5+0.25,-2.5+0.5) (-1.25+1.5,-2.5) node[above] {$04$};
\draw (-1.25+2.0,-0.5) node[above] {$30$};
\draw[fill=gray!50] (-1.25+2.0-0.25,-1.0) rectangle (-1.25+2.0+0.25,-1.0+0.5) (-1.25+2.0,-1.0) node[above] {$01$};
\draw (-1.25+2.0,-1.5) node[above] {$22$};
\draw (-1.25+2.0,-2.0) node[above] {$43$};
\draw (-1.25+2.0,-2.5) node[above] {$14$};
\draw (-1.25+2.5,-0.5) node[above] {$40$};
\draw (-1.25+2.5,-1.0) node[above] {$11$};
\draw (-1.25+2.5,-1.5) node[above] {$32$};
\draw[fill=gray!50] (-1.25+2.5-0.25,-2.0) rectangle (-1.25+2.5+0.25,-2.0+0.5) (-1.25+2.5,-2.0) node[above] {$03$};
\draw (-1.25+2.5,-2.5) node[above] {$24$};
\draw (0.25,-2.5) node[below] {$\pi_2=\left(\begin{smallmatrix} 0& 1& 2& 3& 4\\ 0&3&1&4&2 \end{smallmatrix}\right)$};
\end{tikzpicture}$$

We investigate the distance properties for the polarizing transforms $f(u_1,u_2)=u_1\oplus \pi_i(u_2)$ for $i=1,2$, where $\pi_1=\left(\begin{smallmatrix} 0& 1& 2& 3& 4\\ 0&2&4&1&3 \end{smallmatrix}\right)$ and $\pi_2=\left(\begin{smallmatrix} 0& 1& 2& 3& 4\\ 0&3&1&4&2 \end{smallmatrix}\right)$ by using the tables that the distance properties are the same for $\pi_1$ and $\pi_2$.  Hence, the minimum distance is 
\begin{equation}
d_{\min}=\sqrt{\|s_0-s_1\|^2+\|s_0-s_2\|^2}=2.24\sqrt{E_s}
\label{dmin_eq}
\end{equation}
for 5-PSK signal set, and the distance spectrum is as follows:
\begin{equation}
N(d)=\left\{
\begin{array}{ll}
4   & d=2.24\sqrt{E_s},\\
0   & otherwise.
\end{array}\right.
\end{equation}
for the polarizing transforms $f(u_1,u_2)=u_1\oplus \pi_i(u_2)$ for $i=1,2$.
It is clear to see that these polarizing transforms are equidistant (i.e. $N(d_{min})=q-1$) for 5-PSK signal set, and the minimum distance (\ref{dmin_eq}) of the equidistant transforms is larger than the minimum distance (\ref{dmin_st}) of the standard transform\footnote{Notice that the minimum distances of the class of polarizing transforms introduced in (\ref{polaproc_1}) and the standard transform are the same.}. The distance spectrum upper bound is given for the equidistant transform as follows: 
\begin{eqnarray}
&P_{e}\leq4Q\left(2.24\sqrt{{SNR}/{2}}\right).&
\end{eqnarray} 

Proposition \ref{ww} shows that the distance spectrum upper bound is minimized by the help of an equidistant transform for the $q$-ary signal set. 
Then, we can say that the upper bound of the error performance is improved for the synthetic good channel by using equidistant transforms for a given signal set. To support the claim, we provide simulation results in Fig.~\ref{fig_4} that the error performance of the synthetic good channel is improved for $q=5$ and 5-PSK signal set by using the equidistant transform $f(u_1,u_2)=u_1\oplus \pi_1(u_2)$, where $\pi_1=\left(\begin{smallmatrix} 0& 1& 2& 3& 4\\ 0&2&4&1&3 \end{smallmatrix}\right)$.

We follow the same way to investigate the distance properties of the bad synthetic channel by using the table. 
The analysis of the synthetic bad channel shows that the upper bounds are (almost) the same for any polarizing transform.

\begin{myexample}
Suppose that $(u_1=0,u_2=2)$ and hence $(s_{x_1=2},s_{x_2=2})$ are transmitted. 
For the decision of $u_1$, the following error events can occur:
\begin{enumerate}
\item There are four error events within squared distance $d^2=\|s_0-s_1\|^2$ that $(x_1,x_2)\in\{(1,2),(2,1),(2,3),(3,2)\}$.
\item There are two more error events within squared distance $d^2=2\|s_0-s_1\|^2$ that $(x_1,x_2)\in\{(1,3),(3,1)\}$.
\item There are four more error events within squared distance $d^2=\|s_0-s_2\|^2$ that $(x_1,x_2)\in\{(0,2),(2,0),(2,4),(4,2)\}$.
\item There are eight more error events within squared distance $d^2=\|s_0-s_1\|^2+\|s_0-s_2\|^2$ that $(x_1,x_2)\in\{(1,0),(0,1),(3,0),(0,3),(1,4),(4,1),(3,4),(4,3)\}$.
\item There are two more error events within squared distance $d^2=2\|s_0-s_2\|^2$ that $(x_1,x_2)\in\{(0,4),(4,0)\}$.
\end{enumerate}
\end{myexample}

As such, the minimum distances of the synthetic bad channel are the same,  
\begin{eqnarray}
d_{\min}&=&\|s_i-s_{i+1}\|=1.176\sqrt{E_s},
\end{eqnarray} 
for the standard transform and the equidistant transform for $q$-ary PSK signal set.

The distance spectrum of the standard transform is 
\begin{equation}
N(d)=\left\{
\begin{array}{ll}
4   & d=\|s_0-s_1\|                                       =1.176\sqrt{E_s},\\
2   & d=\sqrt{2}\|s_0-s_1\|                            =1.663\sqrt{E_s},\\
4   & d=\|s_0-s_2\|                                       =1.902\sqrt{E_s},\\
8   & d=\sqrt{\|s_0-s_1\|^2+\|s_0-s_2\|^2}   =2.236\sqrt{E_s},\\
2   & d=\sqrt{2}\|s_0-s_2\|                           =2.690\sqrt{E_s},\\
0   & otherwise.
\end{array}\right.
\end{equation}
for $q=5$ and 5-PSK signal set.

The distance spectrum of the equidistant transform is 
\begin{equation}
N(d)=\left\{
\begin{array}{ll}
4   & d=\|s_0-s_1\|                                       =1.176\sqrt{E_s},\\
4   & d=\sqrt{2}\|s_0-s_1\|                            =1.663\sqrt{E_s},\\
4   & d=\|s_0-s_2\|                                       =1.902\sqrt{E_s},\\
4   & d=\sqrt{\|s_0-s_1\|^2+\|s_0-s_2\|^2}   =2.236\sqrt{E_s},\\
4   & d=\sqrt{2}\|s_0-s_2\|                           =2.690\sqrt{E_s},\\
0   & otherwise.
\end{array}\right.
\end{equation}
for $q=5$ and 5-PSK signal set.
The difference between the upper bounds of the synthetic bad channel is insignificant for the standard transform and the equidistant transform. 
To support the claim, we provide simulation results in Fig.~\ref{fig_5} that the error performances of the synthetic bad channel are (almost) the same for $q=5$ and 5-PSK signal set by using the standard transform and the equidistant transform. 
\end{myexample}

Here, one of the main results in this work is: \textsl{the equidistant transforms for a given $q$-ary signal set provide superior synthetic good channel and (almost) the same synthetic bad channel, and hence, the performance of the error correction capability is improved for the block length $N>2$.} 

Here, superior non-binary synthetic good channel has $N(d_{min})=q-1$ error events a a distance $d_{min}$ which is maximized by the equidistant transform $f$.

In other cases, non-binary synthetic good channels have at least one error event at a distance $d<d_{min}$. Notice that the sum of all distances are constant for all cases. And the equidistant transform maximizes the minimum distance. The equidistant channels was first defined by Jelinek.

We will provide some simulation results in Section \ref{S_simulation_result} in order to strengthen this claim.  
\pagebreak

\begin{figure}[hp]
\centering
\includegraphics[width=0.6\textwidth]{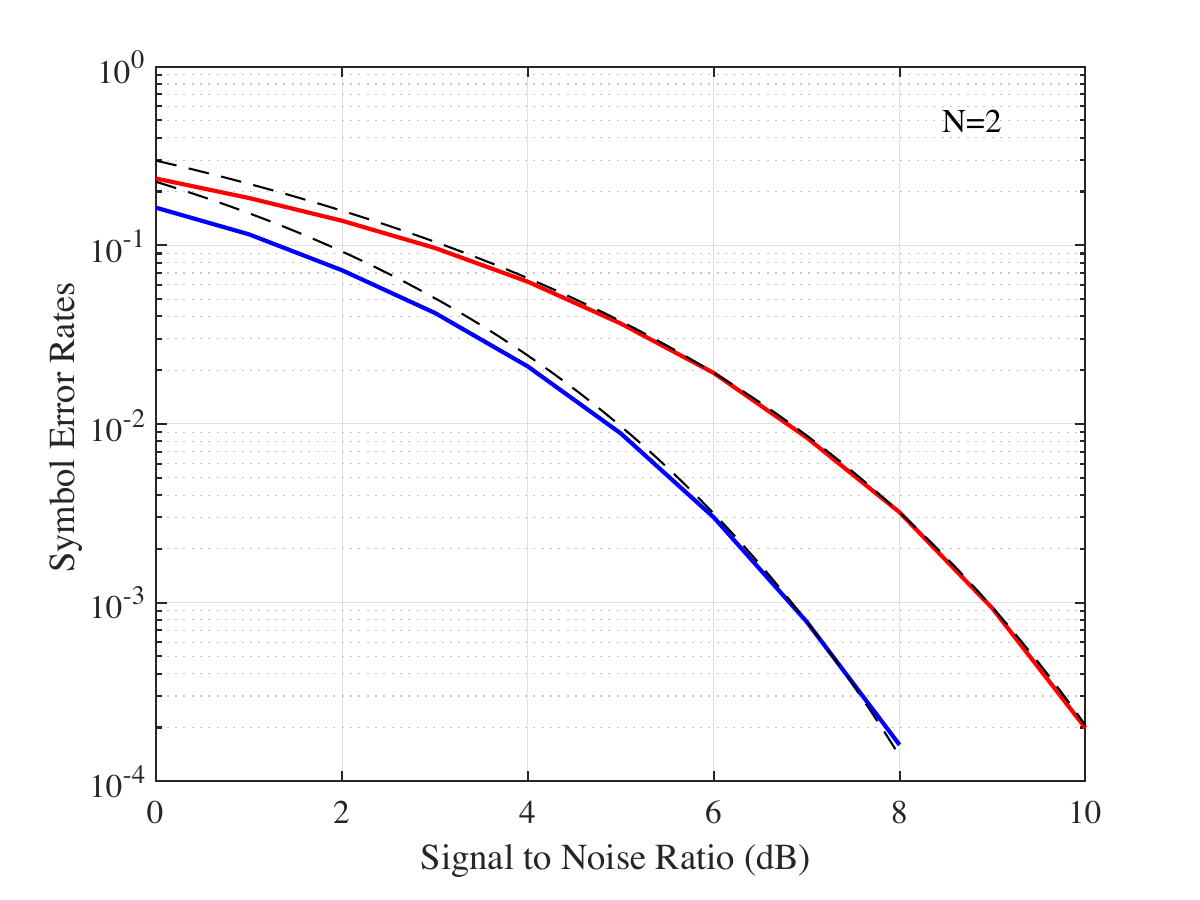}
\caption{Distance spectrum upper bounds (dashed) and simulation results for the equidistant transform (blue) and the standard transform (red) of the synthetic good channel for $q=5$ and 5-PSK signal set.}
\label{fig_4}
\end{figure} 

\begin{figure}[hp]
\centering
\includegraphics[width=0.6\textwidth]{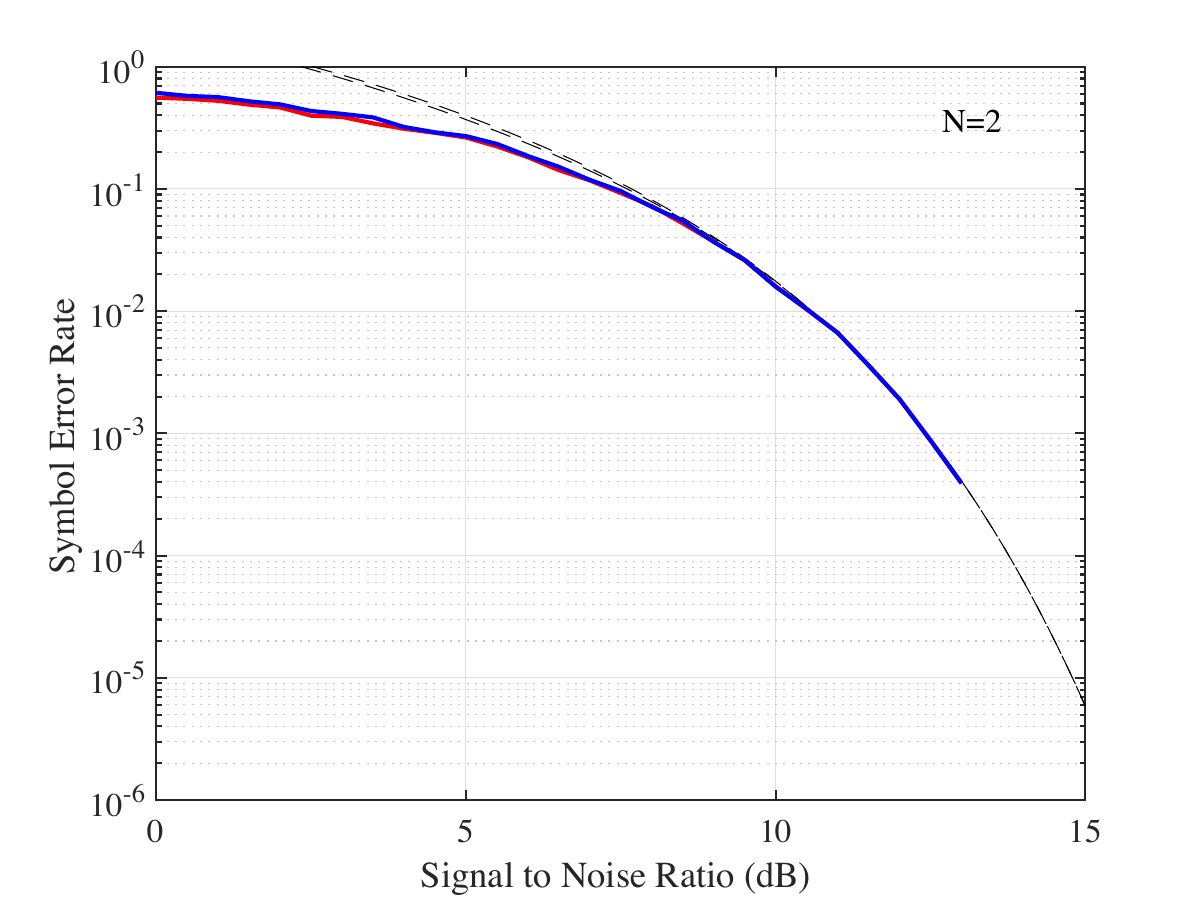}
\caption{Distance spectrum upper bounds (dashed) and simulation results for the equidistant transform (blue) and the standard transform (red) of the synthetic bad channel for $q=5$ and 5-PSK signal set.}
\label{fig_5}
\end{figure}

\pagebreak
\begin{myexample}[A Polarizing Transform for $q=4$]
We applied the procedure for $q=4$ and 4-PSK signal set to design a polarizing transform.
The minimum distance is $\sqrt{2}\|s_0-s_1\|$, and it is the same for all possible polarization transforms of type $f(u_1,u_2)=u_1\oplus \pi(u_2)$.
In this case, the procedure optimizes the distance spectrum $N(d)$ for 4-PSK signal set to improve the upper bound in (\ref{dsb}). 
The first term of the distance spectrum, $N(d_{min})$ is known as the \textsl{kissing number}. 
For this purpose, it minimizes the kissing number by using a particular polarization transform.
The distance spectrum of the standard transform is  
\begin{equation}
N(d)=\left\{
\begin{array}{ll}
2   & d=2\sqrt{E_s},\\
1   & d=\sqrt{\|s_0-s_1\|^2+\|s_0-s_2\|^2}=2.45\sqrt{E_s},\\
0   & otherwise.
\end{array}\right.
\end{equation}  
for $q=4$ and 4-PSK signal set.  
  
The distance spectrum of the optimized transform by using $\pi=\left(\begin{smallmatrix} 0& 1& 2& 3\\ 0&2&1&3 \end{smallmatrix}\right)$ is
\begin{equation}
N(d)=\left\{
\begin{array}{ll}
1   & d=2\sqrt{E_s},\\
2   & d=\sqrt{\|s_0-s_1\|^2+\|s_0-s_2\|^2}=2.45\sqrt{E_s},\\
0   & otherwise.
\end{array}\right.
\end{equation}
for $q=4$ and 4-PSK signal set.
Hence, the distance spectrum upper bound of the standard transform is
\begin{eqnarray}
&P_{e}\leq2Q\left(2\sqrt{\frac{SNR}{2}}\right)+Q\left(2.45\sqrt{\frac{SNR}{2}}\right),&
\label{pe0}
\end{eqnarray} 
and it is reduced to the following upper bound by using the optimized transform for 4-PSK signal set.
\begin{eqnarray}
&P_{e}\leq Q\left(2\sqrt{\frac{SNR}{2}}\right)+2Q\left(2.45\sqrt{\frac{SNR}{2}}\right)&
\label{pe1}
\end{eqnarray} 
It is easy to notice that (\ref{pe0}) is greater than (\ref{pe1}). 
The performance of this method is compared with the set partitioned binary polar codes for multilevel coding in Section \ref{S_simulation_result}.

We also notice that the \textsl{equidistant transform} does not exist for $q=4$ and 4-PSK signal set. In the next section, we will propose a new signal set for $q=4$ in 2-dimensions to design the equidistant transform with the distance spectrum $N({d_{min}})=q-1=3$, and the minimum distance $d_{min}=\sqrt{\frac{2}{3}\sum_{k=1}^{3}\|s_k-s_0\|^2}=2.31$. 
\end{myexample}

\begin{myexample}[Equidistant Polarizing Transforms for q = 3]
To design the equidistant transform, $q=3$ and 3-PSK signal set is the special case by the help of its following property:
$$\|s_j-s_j'\|=\sqrt{3E_s}$$ 
for all $j\neq j'$. The distance spectrum is $N(d_{min})=2$, and the minimum distance is $d_{min}=2.449\sqrt{E_s}$ for all possible polarization transforms, and hence, all polarizing transforms are equidistant for 3-PSK signal set. 
\end{myexample}

Notice that $3$-PSK is only two-dimensional signal set for $q>2$ with constant  $\|s_j-s_j'\|$ for all $j\neq j'$. It is possible to find signal sets with constant distances for $q>3$ in a higher dimension. For that case spectral efficiency is reduced for a given fixed $\lfloor K\log_2 q \rfloor$ number of information bit. In this study, we consider signal sets in 2-dimensions for the non-binary polar coding.

\pagebreak
\begin{myexample}[Almost-Equidistant Transform for $q=8$]
We applied the procedure for $q=8$ and 8-PSK signal set to design a polarizing transform. There does not exist an equidistant transform of type $f(u_1,u_2)=u_1\oplus \pi(u_2)$.
Thanks to the following geometric property of 8-PSK signal set, we can design \textsl{almost}-equidistant transform of type $f(u_1,u_2)=u_1\oplus \pi(u_2)$, where $\pi=\left(\begin{smallmatrix} 0& 1& 2& 3& 4& 5& 6& 7\\ 0&3&6&1&4&7&2&5 \end{smallmatrix}\right)$  by using the procedure for $q=8$.
$${\|s_0-s_1\|^2+\|s_0-s_3\|^2}={2}\|s_0-s_2\|^2=\|s_0-s_4\|^2$$
This geometric property is depicted in Fig.~\ref{Fig_8psk}.

\begin{figure}[hp]
\centering
\begin{tikzpicture}[thick,scale=1.5*1.5*1.4*3.0*0.35, every node/.style={scale=1.1*1.5*1.5*1.25*0.5}]
\draw  (-1.2,0)  -- (1.2,0) ;
\draw  (0,-1.2)  -- (0,1.2) ;
\begin{scope}[very thin]
\draw[<->]  (1,0)  -- (1/1.414213562373095,1/1.414213562373095) ;
\draw[<->]  (1,0)  -- (-1/1.414213562373095,1/1.414213562373095) ;
\draw[<->]  (-1,0)  -- (-1/1.414213562373095,1/1.414213562373095) ;
\draw[<->]  (1,0)  -- (0,1) ;
\draw[<->]  (-1,0)  -- (0,1) ;
\end{scope}
\begin{scope}[very thin,dashed]
\draw (0,0) circle (1.0cm);
\end{scope}
\draw[fill=gray!160!white] (1,0) circle (0.05cm) node[above right] {$s_0$} node[below right] {$$};
\draw[fill=gray!160!white] (1/1.414213562373095,1/1.414213562373095) circle (0.05cm) node[right] {$s_1$} node[ right] {$$};
\draw (0.5+0.5/1.414213562373095,0.5/1.414213562373095)  node[] {$\approx$};
\draw (-0.5-0.5/1.414213562373095,0.5/1.414213562373095)  node[] {$\approx$};
\draw[fill=gray!160!white] (0,1) circle (0.05cm) node[above right] {$s_2$};
\draw (0.00025,0.9995)  node[below] {\begin{tiny}90\textdegree \end{tiny}};
\draw[fill=gray!160!white] (-1/1.414213562373095,1/1.414213562373095) circle (0.05cm) node[left] {$s_3$};
\draw (-1/1.414213562373095+0.05,1/1.414213562373095) node[below] {\begin{tiny}90\textdegree \end{tiny}};
\draw[fill=gray!160!white] (-1.0,0.0) circle (0.05cm) node[below left] {$s_4$} node[above left] {$$};
\draw[fill=gray!160!white] (-1/1.414213562373095,-1/1.414213562373095) circle (0.05cm) node[left] {$s_5$} node[ right] {$$};
\draw[fill=gray!160!white] (0,-1) circle (0.05cm) node[below left] {$s_6$} node[below left] {$$};
\draw[fill=gray!160!white] (1/1.414213562373095,-1/1.414213562373095) circle (0.05cm) node[right] {$s_7$} node[ right] {$$};
\end{tikzpicture}
\caption{The geometric property of $8$-PSK signal set for almost-equidistant transform.}
\label{Fig_8psk}
\end{figure}
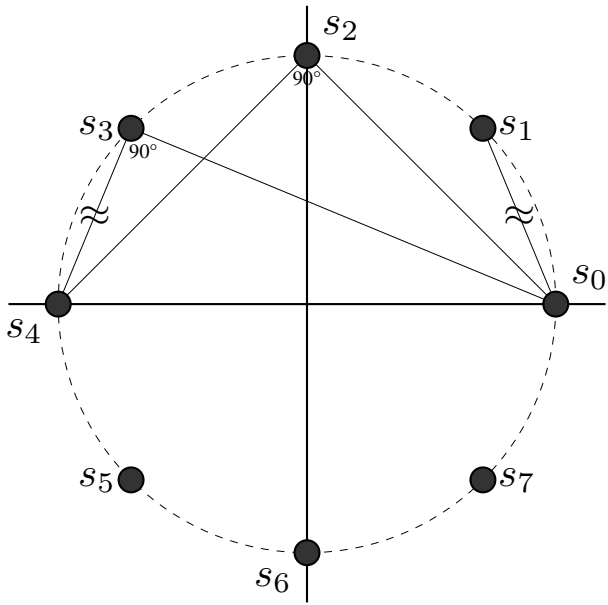

The minimum distance of the almost-equidistant transform is $d_{min}=2\sqrt{E_s}$.
The distance spectrum of the almost-equidistant transform is
\begin{equation}
N(d)=\left\{
\begin{array}{ll}
6   & d=2\sqrt{E_s},\\
1   & d=\sqrt{\|s_0-s_4\|^2+\|s_0-s_4\|^2}=2.83\sqrt{E_s},\\
0   & otherwise.
\end{array}\right.
\end{equation}
for $q=8$ and 8-PSK signal set.
Hence, the distance spectrum upper bound of the almost-equidistant transform is
\begin{eqnarray}
&P_{e}\leq6Q\left(2\sqrt{\frac{SNR}{2}}\right)+Q\left(2.83\sqrt{\frac{SNR}{2}}\right).&
\end{eqnarray} 
If an equidistant transform existed for 8-PSK signal set, the minimum distance would be $d_{min}=2.14\sqrt{E_s}$, and the upper bound would be  
\begin{eqnarray}
&P_{e}\leq7Q\left(2.14\sqrt{\frac{SNR}{2}}\right).&
\end{eqnarray} 
The proposed transform has also (almost) the same with the equidistant upper bound for 8-PSK signal set that can be seen in the Appendix.
\end{myexample}

Notice that the proposed almost equidistant transform with $\pi=\left(\begin{smallmatrix} 0& 1& 2& 3& 4& 5& 6& 7\\ 0&3&6&1&4&7&2&5 \end{smallmatrix}\right)$ has the same anomaly which is defined by \c{S}a\c{s}o\u{g}lu in \cite{sasoglu_polar_2012} to describe failure of the polarization.   
The group $(\mathcal{X},f)$ has a proper nontrivial subgroup. There exists a set  $\mathcal{S}=\{0,4\}$ where $\mathcal{S} \subset \mathcal{X}$ with $|\mathcal{S}| > 1$ such that $(\mathcal{S},f)$ is a group.

It is interesting to notice that with the help of the nearly equidistant characteristic that the proposed transform has, the negative effect on the polarization of the mentioned anomaly can be removed. 
The two-level polarization provided by the almost equidistant transform and the distorted polarization obtained by the standard transform are shown in Fig.~\ref{Fig_2}. Moreover, the speed of polarization obtained by the almost equidistant transform is higher than the polarization obtained by the the non-anomaly transform defined in (\ref{polaproc_1}).
We investigate in more detail the effects of the equidistant property obtained by transform designs on polarization in the Section \ref{S_speed_polarization_exp}.
  
The performance of this method is compared with the non-binary polar code with transform defined in (\ref{polaproc_1}) in Section \ref{S_simulation_result}. We showed that there is a significant difference in performance.

To summarize this section, we provide the distance properties of the proposed and standard transforms for $q$-ary PSK signalling in Table~\ref{table_dist}. 

In the next section, we will describe a new signal set which make possible to define an equidistant transform for $q=4$.

\newpage
\begin{table}[hp]
\centering
\caption{The Distance Properties of Transforms for $q$-ary PSK signal set with $E_s=1 Joule/2-dimensions$} 
\label{table_distance} 
\begin{IEEEeqnarraybox}[\IEEEeqnarraystrutmode\IEEEeqnarraystrutsizeadd{2pt}{0pt}]{x/r/Vx/r/v/r/v/r/v/r/x} \IEEEeqnarraydblrulerowcut\\ 
&&&&\IEEEeqnarraymulticol{3}{t}{Distance Properties of $f:L_q$}&&\IEEEeqnarraymulticol{3}{t}{Distance Properties of $\left[\begin{smallmatrix}1&0\\ 1&1\end{smallmatrix}\right]$}&\\ 
&\hfill\raisebox{-3pt}[0pt][0pt]{$q$}\hfill&&\IEEEeqnarraymulticol{9}{h}{} 
\IEEEeqnarraystrutsize{0pt}{0pt}\\ 
&&&&\hfill d_{min}\hfill&&\hfill
N(d)\hfill&&\hfill d_{min}\hfill&&\hfill
N(d)\hfill&\IEEEeqnarraystrutsizeadd{0pt}{2pt}\\
\IEEEeqnarraydblrulerowcut\\
&3&&& 2.449&&    2&& 2.449&&2&\\
&4&&& 2.000&& 1,2&& 2.000&&2,1&\\
&5&&& 2.236&&    4&& 1.663&&2,2&\\
&6&&& 2.000&& -&& 1.414&&2,2,1&\\
&8&&& 2.000&& 6,1&& 1.082&&2,2,2,1&\\
\IEEEeqnarraydblrulerowcut\\ 
\end{IEEEeqnarraybox}
\label{table_dist}
\end{table}

\begin{center}
\begin{tikzpicture}[scale=0.95, every node/.style={scale=0.95}]
\draw(-3.5,-0.75) node[] {$L3:\pi=\left(\begin{smallmatrix} 0& 1& 2\\ 0&2&1 \end{smallmatrix}\right)$};
\draw  (-1.0,0)  -- (1.0-0.5,0) ;
\draw  (-1.0,-0.5)  -- (1.0-0.5,-0.5) ;
\draw  (-1.0,-1.0)  -- (1.0-0.5,-1.0) ;
\draw  (-1.0,-1.5)  -- (1.0-0.5,-1.5) ;
\draw  (-1.0, 0.0)  -- (-1.0,-2.0+0.5) ;
\draw  (-0.5, 0.0)  -- (-0.5,-2.0+0.5) ;
\draw  (-0.0, 0.0)  -- (-0.0,-2.0+0.5) ;
\draw  (0.5, 0.0)  -- (0.5,-2.0+0.5) ;
\draw[fill=gray!50] (-1.25+0.5-0.25,-0.5) rectangle (-1.25+0.5+0.5-0.25,-0.5+0.5) (-1.25+0.5+0.25-0.25,-0.5) node[above]  {$00$};%
\draw (-1.25+0.5,-1.0) node[above] {$11$};
\draw (-1.25+0.5,-1.5) node[above] {$22$};
\draw (-1.25+1.0,-0.5) node[above] {$10$};
\draw (-1.25+1.0,-1.0) node[above] {$21$};
\draw[fill=gray!50] (-1.25+1.0-0.25,-1.5) rectangle (-1.25+1.0+0.5-0.25,-1.5+0.5) (-1.25+1.0+0.25-0.25,-1.5) node[above]  {$02$};%
\draw (-1.25+1.5,-0.5) node[above] {$20$};
\draw[fill=gray!50] (-1.25+1.5-0.25,-1.0) rectangle (-1.25+1.5+0.5-0.25,-1.0+0.5) (-1.25+1.5+0.25-0.25,-1.0) node[above]  {$01$};%
\draw (-1.25+1.5,-1.5) node[above] {$12$};
\end{tikzpicture}
\end{center}
\begin{center}
\begin{tikzpicture}[scale=0.95, every node/.style={scale=0.95}]
\draw(-3.5,-1.00) node[] {$L4:\pi=\left(\begin{smallmatrix} 0& 1& 2& 3\\ 0&2&1&3 \end{smallmatrix}\right)$};
\draw  (-1.0, 0.0)  -- (1.0, 0.0) ;
\draw  (-1.0,-0.5)  -- (1.0,-0.5) ;
\draw  (-1.0,-1.0)  -- (1.0,-1.0) ;
\draw  (-1.0,-1.5)  -- (1.0,-1.5) ;
\draw  (-1.0,-2.0)  -- (1.0,-2.0) ;
\draw  (-1.0, 0.0)  -- (-1.0,-2.0) ;
\draw  (-0.5, 0.0)  -- (-0.5,-2.0) ;
\draw  (-0.0, 0.0)  -- (-0.0,-2.0) ;
\draw  ( 0.5, 0.0)  -- ( 0.5,-2.0) ;
\draw  ( 1.0, 0.0)  -- ( 1.0,-2.0) ;
\draw[fill=gray!50] (-1.25+0.5-0.25,-0.5) rectangle (-1.25+0.5+0.5-0.25,-0.5+0.5) (-1.25+0.5+0.25-0.25,-0.5) node[above]  {$00$};%
\draw (-1.25+0.5,-1.0) node[above] {$21$};
\draw (-1.25+0.5,-1.5) node[above] {$32$};
\draw (-1.25+0.5,-2.0) node[above] {$13$};
\draw (-1.25+1.0,-0.5) node[above] {$10$};
\draw (-1.25+1.0,-1.0) node[above] {$31$};
\draw[fill=gray!50] (-1.25+1.0-0.25,-1.5) rectangle (-1.25+1.0+0.5-0.25,-1.5+0.5) (-1.25+1.0+0.25-0.25,-1.5) node[above]  {$02$};%
\draw (-1.25+1.0,-2.0) node[above]  {$23$};
\draw (-1.25+1.5,-0.5) node[above] {$20$};
\draw[fill=gray!50] (-1.25+1.5-0.25,-1.0) rectangle (-1.25+1.5+0.5-0.25,-1.0+0.5) (-1.25+1.5+0.25-0.25,-1.0) node[above]   {$01$};%
\draw (-1.25+1.5,-1.5) node[above] {$12$};
\draw (-1.25+1.5,-2.0) node[above] {$33$};
\draw (-1.25+2,-0.5) node[above]    {$30$};
\draw (-1.25+2,-1.0) node[above]     {$11$};
\draw (-1.25+2,-1.5) node[above]    {$22$};
\draw[fill=gray!50] (-1.25+2-0.25,-2.0) rectangle (-1.25+2+0.5-0.25,-2.0+0.5) (-1.25+2+0.25-0.25,-2.0) node[above]    {$03$};%
\end{tikzpicture}
\end{center}
\begin{center}
\begin{tikzpicture}[scale=0.95, every node/.style={scale=0.95}]
\draw(-3.5,-1.25) node[] {$L5:\pi=\left(\begin{smallmatrix} 0& 1& 2& 3& 4\\ 0&3&1&4&2 \end{smallmatrix}\right)$};
\draw  (-1.0, 0.0)  -- (1.5, 0.0) ;
\draw  (-1.0,-0.5)  -- (1.5,-0.5) ;
\draw  (-1.0,-1.0)  -- (1.5,-1.0) ;
\draw  (-1.0,-1.5)  -- (1.5,-1.5) ;
\draw  (-1.0,-2.0)  -- (1.5,-2.0) ;
\draw  (-1.0,-2.5)  -- (1.5,-2.5) ;
\draw  (-1.0, 0.0)  -- (-1.0,-2.5) ;
\draw  (-0.5, 0.0)  -- (-0.5,-2.5) ;
\draw  (-0.0, 0.0)  -- (-0.0,-2.5) ;
\draw  ( 0.5, 0.0)  -- ( 0.5,-2.5) ;
\draw  ( 1.0, 0.0)  -- ( 1.0,-2.5) ;
\draw  ( 1.5, 0.0)  -- ( 1.5,-2.5) ;
\draw[fill=gray!50] (-1.25+0.5-0.25,-0.5) rectangle (-1.25+0.5+0.5-0.25,-0.5+0.5) (-1.25+0.5+0.25-0.25,-0.5) node[above]  {$00$};%
\draw (-1.25+0.5,-1.0) node[above] {$21$};
\draw (-1.25+0.5,-1.5) node[above] {$42$};
\draw (-1.25+0.5,-2.0) node[above] {$13$};
\draw (-1.25+0.5,-2.5) node[above] {$34$};
\draw (-1.25+1.0,-0.5) node[above] {$10$};
\draw (-1.25+1.0,-1.0) node[above] {$31$};
\draw[fill=gray!50] (-1.25+1.0-0.25,-1.5) rectangle (-1.25+1.0+0.5-0.25,-1.5+0.5) (-1.25+1.0+0.25-0.25,-1.5) node[above]  {$02$};%
\draw (-1.25+1.0,-2.0) node[above]  {$23$};
\draw (-1.25+1.0,-2.5) node[above]  {$44$};
\draw (-1.25+1.5,-0.5) node[above] {$20$};
\draw (-1.25+1.5,-1.0) node[above]  {$41$};
\draw (-1.25+1.5,-1.5) node[above] {$12$};
\draw (-1.25+1.5,-2.0) node[above] {$33$};
\draw[fill=gray!50] (-1.25+1.5-0.25,-2.5) rectangle (-1.25+1.5+0.5-0.25,-2.5+0.5) (-1.25+1.5+0.25-0.25,-2.5) node[above] {$04$};%
\draw (-1.25+2,-0.5) node[above]    {$30$};
\draw[fill=gray!50] (-1.25+2-0.25,-1.0) rectangle (-1.25+2+0.5-0.25,-1.0+0.5) (-1.25+2+0.25-0.25,-1.0) node[above]     {$01$};%
\draw (-1.25+2,-1.5) node[above]    {$22$};
\draw (-1.25+2,-2.0) node[above]    {$43$};
\draw (-1.25+2,-2.5) node[above]    {$14$};
\draw (-1.25+2.5,-0.5) node[above] {$40$};
\draw (-1.25+2.5,-1.0) node[above]  {$11$};
\draw (-1.25+2.5,-1.5) node[above] {$32$};
\draw[fill=gray!50] (-1.25+2.5-0.25,-2.0) rectangle (-1.25+2.5+0.5-0.25,-2.0+0.5) (-1.25+2.5+0.25-0.25,-2.0) node[above] {$03$};%
\draw (-1.25+2.5,-2.5) node[above] {$24$};
\end{tikzpicture}
\end{center}
\begin{center}
\begin{tikzpicture}[scale=0.95, every node/.style={scale=0.95}]
\draw(-3.5,-1.5) node[] {$L6:\pi=\left(\begin{smallmatrix} 0& 1& 2& 3& 4& 5\\ 0&4&1&3&5&2 \end{smallmatrix}\right)$};
\draw  (-1.0, 0.0)  -- (2.0, 0.0) ;
\draw  (-1.0,-0.5)  -- (2.0,-0.5) ;
\draw  (-1.0,-1.0)  -- (2.0,-1.0) ;
\draw  (-1.0,-1.5)  -- (2.0,-1.5) ;
\draw  (-1.0,-2.0)  -- (2.0,-2.0) ;
\draw  (-1.0,-2.5)  -- (2.0,-2.5) ;
\draw  (-1.0,-3.0)  -- (2.0,-3.0) ;
\draw  (-1.0, 0.0)  -- (-1.0,-3.0) ;
\draw  (-0.5, 0.0)  -- (-0.5,-3.0) ;
\draw  (-0.0, 0.0)  -- (-0.0,-3.0) ;
\draw  ( 0.5, 0.0)  -- ( 0.5,-3.0) ;
\draw  ( 1.0, 0.0)  -- ( 1.0,-3.0) ;
\draw  ( 1.5, 0.0)  -- ( 1.5,-3.0) ;
\draw  ( 2.0, 0.0)  -- ( 2.0,-3.0) ;
\draw[fill=gray!50] (-1.25+0.5-0.25,-0.5) rectangle (-1.25+0.5+0.5-0.25,-0.5+0.5) (-1.25+0.5+0.25-0.25,-0.5) node[above]  {$00$};%
\draw (-1.25+0.5,-1.0) node[above] {$21$};
\draw (-1.25+0.5,-1.5) node[above] {$52$};
\draw (-1.25+0.5,-2.0) node[above] {$33$};
\draw (-1.25+0.5,-2.5) node[above] {$14$};
\draw (-1.25+0.5,-3.0) node[above] {$45$};%
\draw (-1.25+1.0,-0.5) node[above] {$10$};
\draw (-1.25+1.0,-1.0) node[above] {$31$};
\draw[fill=gray!50] (-1.25+1.0-0.25,-1.5) rectangle (-1.25+1.0+0.5-0.25,-1.5+0.5) (-1.25+1.0+0.25-0.25,-1.5) node[above]  {$02$};%
\draw[fill=gray!0] (-1.25+1.0-0.25,-2.0) rectangle (-1.25+1.0+0.5-0.25,-2.0+0.5) (-1.25+1.0+0.25-0.25,-2.0) node[above]  {$43$};
\draw (-1.25+1.0,-2.5) node[above]  {$24$};
\draw (-1.25+1.0,-3.0) node[above]  {$55$};%
\draw (-1.25+1.5,-0.5) node[above] {$20$};
\draw (-1.25+1.5,-1.0) node[above]  {$41$};
\draw[fill=gray!0] (-1.25+1.5-0.25,-1.5) rectangle (-1.25+1.5+0.5-0.25,-1.5+0.5) (-1.25+1.5+0.25-0.25,-1.5) node[above] {$12$};
\draw (-1.25+1.5,-2.0) node[above] {$53$};
\draw (-1.25+1.5,-2.5) node[above] {$34$};
\draw[fill=gray!50] (-1.25+1.5-0.25,-3.0) rectangle (-1.25+1.5+0.5-0.25,-3.0+0.5) (-1.25+1.5+0.25-0.25,-3.0) node[above] {$05$};%
\draw (-1.25+2,-0.5) node[above]    {$30$};
\draw (-1.25+2,-1.0) node[above]     {$51$};
\draw (-1.25+2,-1.5) node[above]    {$22$};
\draw[fill=gray!50] (-1.25+2-0.25,-2.0) rectangle (-1.25+2+0.5-0.25,-2.0+0.5) (-1.25+2+0.25-0.25,-2.0) node[above]    {$03$};%
\draw (-1.25+2,-2.5) node[above]    {$44$};
\draw[fill=gray!0] (-1.25+2-0.25,-3.0) rectangle (-1.25+2+0.5-0.25,-3.0+0.5) (-1.25+2+0.25-0.25,-3.0) node[above]    {$15$};
\draw (-1.25+2.5,-0.5) node[above] {$40$};
\draw[fill=gray!50] (-1.25+2.5-0.25,-1.0) rectangle (-1.25+2.5+0.5-0.25,-1.0+0.5) (-1.25+2.5+0.25-0.25,-1.0) node[above]  {$01$};%
\draw (-1.25+2.5,-1.5) node[above] {$32$};
\draw (-1.25+2.5,-2.0) node[above] {$13$};
\draw[fill=gray!0] (-1.25+2.5-0.25,-2.5) rectangle (-1.25+2.5+0.5-0.25,-2.5+0.5) (-1.25+2.5+0.25-0.25,-2.5) node[above] {$54$};
\draw (-1.25+2.5,-3.0) node[above] {$25$};%
\draw (-1.25+3,-0.5) node[above]    {$50$};
\draw[fill=gray!0] (-1.25+3-0.25,-1.0) rectangle (-1.25+3+0.5-0.25,-1.0+0.5) (-1.25+3+0.25-0.25,-1.0) node[above]     {$11$};
\draw (-1.25+3,-1.5) node[above]    {$42$};
\draw (-1.25+3,-2.0) node[above]    {$23$};
\draw[fill=gray!50] (-1.25+3-0.25,-2.5) rectangle (-1.25+3+0.5-0.25,-2.5+0.5) (-1.25+3+0.25-0.25,-2.5) node[above]    {$04$};%
\draw (-1.25+3,-3.0) node[above]    {$35$};
\end{tikzpicture}
\end{center}
\begin{center}
\begin{tikzpicture}[scale=0.95, every node/.style={scale=0.95}]
\draw(-3.5,-1.75) node[] {$L7:\pi=\left(\begin{smallmatrix} 0& 1& 2& 3& 4& 5& 6\\ 0&3&5&1&6&2&4 \end{smallmatrix}\right)$};
\draw  (-1.0, 0.0)  -- (2.5, 0.0) ;
\draw  (-1.0,-0.5)  -- (2.5,-0.5) ;
\draw  (-1.0,-1.0)  -- (2.5,-1.0) ;
\draw  (-1.0,-1.5)  -- (2.5,-1.5) ;
\draw  (-1.0,-2.0)  -- (2.5,-2.0) ;
\draw  (-1.0,-2.5)  -- (2.5,-2.5) ;
\draw  (-1.0,-3.0)  -- (2.5,-3.0) ;
\draw  (-1.0,-3.5)  -- (2.5,-3.5) ;
\draw  (-1.0, 0.0)  -- (-1.0,-3.5) ;
\draw  (-0.5, 0.0)  -- (-0.5,-3.5) ;
\draw  (-0.0, 0.0)  -- (-0.0,-3.5) ;
\draw  ( 0.5, 0.0)  -- ( 0.5,-3.5) ;
\draw  ( 1.0, 0.0)  -- ( 1.0,-3.5) ;
\draw  ( 1.5, 0.0)  -- ( 1.5,-3.5) ;
\draw  ( 2.0, 0.0)  -- ( 2.0,-3.5) ;
\draw  ( 2.5, 0.0)  -- ( 2.5,-3.5) ;
\draw[fill=gray!50] (-1.25+0.5-0.25,-0.5) rectangle (-1.25+0.5+0.5-0.25,-0.5+0.5) (-1.25+0.5+0.25-0.25,-0.5) node[above]  {$00$};%
\draw (-1.25+0.5,-1.0) node[above] {$41$};
\draw (-1.25+0.5,-1.5) node[above] {$22$};
\draw (-1.25+0.5,-2.0) node[above] {$63$};
\draw (-1.25+0.5,-2.5) node[above] {$14$};
\draw (-1.25+0.5,-3.0) node[above] {$55$};
\draw (-1.25+0.5,-3.5) node[above] {$36$};%
\draw (-1.25+1.0,-0.5) node[above] {$10$};
\draw (-1.25+1.0,-1.0) node[above] {$51$};
\draw[fill=gray!0] (-1.25+1.0-0.25,-1.5) rectangle (-1.25+1.0+0.5-0.25,-1.5+0.5) (-1.25+1.0+0.25-0.25,-1.5) node[above]  {$32$};%
\draw[fill=gray!50] (-1.25+1.0-0.25,-2.0) rectangle (-1.25+1.0+0.5-0.25,-2.0+0.5) (-1.25+1.0-0.25+0.25,-2.0) node[above]  {$03$};
\draw (-1.25+1.0,-2.5) node[above]  {$24$};
\draw (-1.25+1.0,-3.0) node[above]  {$65$};
\draw (-1.25+1.0,-3.5) node[above]  {$46$};%
\draw (-1.25+1.5,-0.5) node[above] {$20$};
\draw (-1.25+1.5,-1.0) node[above]  {$61$};
\draw (-1.25+1.5,-1.5) node[above] {$42$};
\draw (-1.25+1.5,-2.0) node[above] {$13$};
\draw[fill=gray!0] (-1.25+1.5-0.25,-2.5) rectangle (-1.25+1.5+0.5-0.25,-2.5+0.5) (-1.25+1.5+0.25-0.25,-2.5) node[above] {$34$};%
\draw[fill=gray!50] (-1.25+1.5-0.25,-3.0) rectangle (-1.25+1.5-0.25+0.5,-3.0+0.5) (-1.25+1.5-0.25+0.25,-3.0)  node[above] {$05$};
\draw (-1.25+1.5,-3.5) node[above] {$56$};%
\draw (-1.25+2,-0.5) node[above]    {$30$};
\draw[fill=gray!50] (-1.25+2-0.25,-1.0) rectangle (-1.25+2-0.25+0.5,-1.0+0.5) (-1.25+2-0.25+0.25,-1.0) node[above]     {$01$};
\draw (-1.25+2,-1.5) node[above]    {$52$};
\draw (-1.25+2,-2.0) node[above]    {$23$};
\draw (-1.25+2,-2.5) node[above]    {$44$};
\draw (-1.25+2,-3.0) node[above]    {$15$};
\draw[fill=gray!0] (-1.25+2-0.25,-3.5) rectangle (-1.25+2+0.5-0.25,-3.5+0.5) (-1.25+2+0.25-0.25,-3.5) node[above]    {$66$};%
\draw (-1.25+2.5,-0.5) node[above] {$40$};
\draw[fill=gray!0] (-1.25+2.5-0.25,-1.0) rectangle (-1.25+2.5+0.5-0.25,-1.0+0.5) (-1.25+2.5+0.25-0.25,-1.0) node[above]  {$11$};%
\draw (-1.25+2.5,-1.5) node[above] {$62$};
\draw (-1.25+2.5,-2.0) node[above] {$33$};
\draw (-1.25+2.5,-2.5) node[above] {$54$};
\draw (-1.25+2.5,-3.0) node[above] {$25$};
\draw[fill=gray!50] (-1.25+2.5-0.25,-3.5) rectangle (-1.25+2.5-0.25+0.5,-3.5+0.5) (-1.25+2.5-0.25+0.25,-3.5) node[above] {$06$};%
\draw (-1.25+3,-0.5) node[above]    {$50$};
\draw (-1.25+3,-1.0) node[above]     {$21$};
\draw[fill=gray!50] (-1.25+3-0.25,-1.5) rectangle (-1.25+3-0.25+0.5,-1.5+0.5) (-1.25+3-0.25+0.25,-1.5) node[above]    {$02$};
\draw[fill=gray!0] (-1.25+3-0.25,-2.0) rectangle (-1.25+3+0.5-0.25,-2.0+0.5) (-1.25+3+0.25-0.25,-2.0) node[above]    {$43$};%
\draw (-1.25+3,-2.5) node[above]    {$64$};
\draw (-1.25+3,-3.0) node[above]    {$35$};
\draw (-1.25+3,-3.5) node[above]    {$16$};
\draw (-1.25+3.5,-0.5) node[above] {$60$};
\draw (-1.25+3.5,-1.0) node[above]  {$31$};
\draw (-1.25+3.5,-1.5) node[above] {$12$};
\draw (-1.25+3.5,-2.0) node[above] {$53$};
\draw[fill=gray!50] (-1.25+3.5-0.25,-2.5) rectangle (-1.25+3.5-0.25+0.5,-2.5+0.5) (-1.25+3.5-0.25+0.25,-2.5) node[above] {$04$};
\draw[fill=gray!0] (-1.25+3.5-0.25,-3.0) rectangle (-1.25+3.5+0.5-0.25,-3.0+0.5) (-1.25+3.5+0.25-0.25,-3.0) node[above] {$45$};%
\draw (-1.25+3.5,-3.5) node[above] {$26$};
\end{tikzpicture}
\end{center}
\begin{center}
\begin{tikzpicture}[scale=0.95, every node/.style={scale=0.95}]
\draw(-3.5,-2.00) node[] {$L8:\pi=\left(\begin{smallmatrix} 0& 1& 2& 3& 4& 5& 6& 7\\ 0&3&6&1&4&7&2&5 \end{smallmatrix}\right)$};
\draw  (-1.0, 0.0)  -- (3, 0.0) ;
\draw  (-1.0,-0.5)  -- (3,-0.5) ;
\draw  (-1.0,-1.0)  -- (3,-1.0) ;
\draw  (-1.0,-1.5)  -- (3,-1.5) ;
\draw  (-1.0,-2.0)  -- (3,-2.0) ;
\draw  (-1.0,-2.5)  -- (3,-2.5) ;
\draw  (-1.0,-3.0)  -- (3,-3.0) ;
\draw  (-1.0,-3.5)  -- (3,-3.5) ;
\draw  (-1.0,-4.0)  -- (3,-4.0) ;
\draw  (-1.0, 0.0)  -- (-1.0,-4) ;
\draw  (-0.5, 0.0)  -- (-0.5,-4) ;
\draw  (-0.0, 0.0)  -- (-0.0,-4) ;
\draw  ( 0.5, 0.0)  -- ( 0.5,-4) ;
\draw  ( 1.0, 0.0)  -- ( 1.0,-4) ;
\draw  ( 1.5, 0.0)  -- ( 1.5,-4) ;
\draw  ( 2.0, 0.0)  -- ( 2.0,-4) ;
\draw  ( 2.5, 0.0)  -- ( 2.5,-4) ;
\draw  ( 3.0, 0.0)  -- ( 3.0,-4) ;
\draw[fill=gray!50] (-1.25+0.5-0.25,-0.5) rectangle (-1.25+0.5-0.25+0.5,-0.5+0.5) (-1.25+0.5+0.25-0.25,-0.5) node[above]  {$00$};%
\draw (-1.25+0.5,-1.0) node[above] {$51$};
\draw (-1.25+0.5,-1.5) node[above] {$22$};
\draw (-1.25+0.5,-2.0) node[above] {$73$};
\draw (-1.25+0.5,-2.5) node[above] {$44$};
\draw (-1.25+0.5,-3.0) node[above] {$15$};
\draw (-1.25+0.5,-3.5) node[above] {$66$};
\draw (-1.25+0.5,-4.0) node[above] {$37$};
\draw (-1.25+1.0,-0.5) node[above] {$10$};
\draw (-1.25+1.0,-1.0) node[above] {$61$};
\draw (-1.25+1.0,-1.5) node[above]  {$32$};
\draw[fill=gray!50] (-1.25+1.0-0.25,-2.0) rectangle (-1.25+1.0-0.25+0.5,-2.0+0.5) (-1.25+1.0+0.25-0.25,-2.0) node[above]  {$03$};%
\draw (-1.25+1.0,-2.5) node[above]  {$54$};
\draw (-1.25+1.0,-3.0) node[above]  {$25$};
\draw (-1.25+1.0,-3.5) node[above]  {$76$};
\draw (-1.25+1.0,-4.0) node[above]  {$47$};
\draw (-1.25+1.5,-0.5) node[above] {$20$};
\draw (-1.25+1.5,-1.0) node[above]  {$71$};
\draw (-1.25+1.5,-1.5) node[above] {$42$};
\draw (-1.25+1.5,-2.0) node[above] {$13$};
\draw (-1.25+1.5,-2.5) node[above] {$64$};
\draw (-1.25+1.5,-3.0) node[above] {$35$};
\draw[fill=gray!50] (-1.25+1.5-0.25,-3.5) rectangle (-1.25+1.5-0.25+0.5,-3.5+0.5) (-1.25+1.5+0.25-0.25,-3.5) node[above] {$06$};%
\draw (-1.25+1.5,-4.0) node[above] {$57$};
\draw (-1.25+2,-0.5) node[above]    {$30$};
\draw[fill=gray!50] (-1.25+2-0.25,-1.0) rectangle (-1.25+2-0.25+0.5,-1.0+0.5) (-1.25+2+0.25-0.25,-1.0) node[above]     {$01$};%
\draw (-1.25+2,-1.5) node[above]    {$52$};
\draw (-1.25+2,-2.0) node[above]    {$23$};
\draw (-1.25+2,-2.5) node[above]    {$74$};
\draw (-1.25+2,-3.0) node[above]    {$45$};
\draw (-1.25+2,-3.5) node[above]    {$16$};
\draw (-1.25+2,-4.0) node[above]    {$67$};
\draw (-1.25+2.5,-0.5) node[above] {$40$};
\draw (-1.25+2.5,-1.0) node[above]  {$11$};
\draw (-1.25+2.5,-1.5) node[above] {$62$};
\draw (-1.25+2.5,-2.0) node[above] {$33$};
\draw[fill=gray!50] (-1.25+2.5-0.25,-2.5) rectangle (-1.25+2.5-0.25+0.5,-2.5+0.5) (-1.25+2.5+0.25-0.25,-2.5) node[above] {$04$};%
\draw (-1.25+2.5,-3.0) node[above] {$55$};
\draw (-1.25+2.5,-3.5) node[above] {$26$};
\draw (-1.25+2.5,-4.0) node[above] {$77$};
\draw (-1.25+3,-0.5) node[above]    {$50$};
\draw (-1.25+3,-1.0) node[above]     {$21$};
\draw (-1.25+3,-1.5) node[above]    {$72$};
\draw (-1.25+3,-2.0) node[above]    {$43$};
\draw (-1.25+3,-2.5) node[above]    {$14$};
\draw (-1.25+3,-3.0) node[above]    {$65$};
\draw (-1.25+3,-3.5) node[above]    {$36$};
\draw[fill=gray!50] (-1.25+3-0.25,-4.0) rectangle (-1.25+3-0.25+0.5,-4.0+0.5) (-1.25+3+0.25-0.25,-4.0) node[above]    {$07$};%
\draw (-1.25+3.5,-0.5) node[above] {$60$};
\draw (-1.25+3.5,-1.0) node[above]  {$31$};
\draw[fill=gray!50] (-1.25+3.5-0.25,-1.5) rectangle (-1.25+3.5-0.25+0.5,-1.5+0.5) (-1.25+3.5+0.25-0.25,-1.5) node[above] {$02$};%
\draw (-1.25+3.5,-2.0) node[above] {$53$};
\draw (-1.25+3.5,-2.5) node[above] {$24$};
\draw (-1.25+3.5,-3.0) node[above] {$75$};
\draw (-1.25+3.5,-3.5) node[above] {$46$};
\draw (-1.25+3.5,-4.0) node[above] {$17$};
\draw (-1.25+4,-0.5) node[above]    {$70$};
\draw (-1.25+4,-1.0) node[above]     {$41$};
\draw (-1.25+4,-1.5) node[above]    {$12$};
\draw (-1.25+4,-2.0) node[above]    {$63$};
\draw (-1.25+4,-2.5) node[above]    {$34$};
\draw[fill=gray!50] (-1.25+4-0.25,-3.0) rectangle (-1.25+4-0.25+0.5,-3.0+0.5) (-1.25+4+0.25-0.25,-3.0) node[above]    {$05$};%
\draw (-1.25+4,-3.5) node[above]    {$56$};
\draw (-1.25+4,-4.0) node[above]    {$27$};
\end{tikzpicture}
\end{center}

\pagebreak

\section{Signal-Set Design for Nonbinary Equidistance}\label{S_signalset_design}  

\begin{figure}[hp]
\centering
\begin{tikzpicture}[thick,scale=1.5*0.5*1.1*2.5*1.1*3.0*0.35, every node/.style={scale=0.9*1.4*1.25*0.5}]
\draw  (-1.2,0)  -- (1.2,0) ;
\draw  (0,-1.2)  -- (0,1.2) ;
\begin{scope}[very thin]
\draw[<->]  (0,0)  -- (0.866,0.5) ;
\draw  (0.333,0.25) node[above] {$\begin{smallmatrix}\sqrt{E_s}\end{smallmatrix}$};
\end{scope}
\begin{scope}[very thin,dashed]
\draw (0,0) circle (1.0cm);
\end{scope}
\draw[fill=gray!160!white] (1,0) circle (0.05cm) node[above right] {$s_0$} node[below] {$(1,0)$};
\begin{scope}[very thin,dashed]
\draw[fill=gray!10!white]  (0,1) circle (0.05cm);
\draw[fill=gray!10!white]  (0,-1) circle (0.05cm);
\end{scope}
\draw[<-] (1.2*1/3,1.2*0.94280904158) arc (69:87:1.2);
\draw[<-] (-1.2*1/3,-1.2*0.94280904158) arc (69+180:87+180:1.2);
\draw[fill=gray!160!white] (1/3,0.94280904158) circle (0.05cm) node[above] {$s_1$} node[ right] {$\left(\frac{1}{3},\frac{2\sqrt{2}}{3}\right)$};
\draw[fill=gray!160!white] (-1.0,0.0) circle (0.05cm) node[below left] {$s_2$} node[above] {$(-1,0)$};
\draw[fill=gray!160!white] (-1/3,-0.94280904158) circle (0.05cm) node[below] {$s_3$} node[ left] {$\left(-\frac{1}{3},-\frac{2\sqrt{2}}{3}\right)$};
\end{tikzpicture}
\caption{The new signal-set with $E_s=1$ joule/2-dimension for $q=4$ nonbinary equidistance.}
\label{Fig_6}
\end{figure}
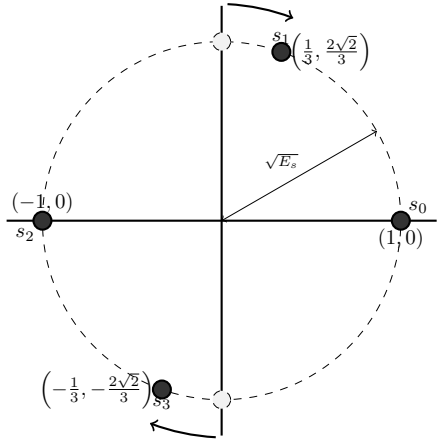

The new signal-set for $q=4$ is obtained by the changing the geometry of the 4-PSK signaling as shown in Fig.~\ref{Fig_6}. 
In this way, an equal rotation in clockwise direction was applied to the $s_1$ and $s_3$. 
The following relation is found between the new Euclidean distances, depending on the amount of rotation as a result of the rotation process. 
$$\|s_0-s_1\|=x\sqrt{E_s},$$ 
$$\|s_0-s_3\|=\sqrt{4-x^2}\sqrt{E_s}.$$
Here, $x$ is a variable that depends on the rotation. The following equation must be satisfied in order to have the equidistant characteristic of the transform given above. 
\begin{equation}
\sqrt{\|s_0-s_1\|^2+\|s_0-s_2\|^2}=\sqrt{\|s_0-s_3\|^2+\|s_0-s_3\|^2}
\label{sigse}
\end{equation} 
For this, $x=\frac{2}{\sqrt{3}}$ is found and the amount of rotation is determined.
As a result of these, PSK-type the new signal-set is depicted in Fig.~\ref{Fig_6}.

The distance properties are optimized by the nonbinary equidistance for the new signal-set for a fixed $u_1=2$ as follows
The minimum distance is $d_{min}=2.309\sqrt{E_s}$
The distance spectrum is $N(d_{min})=3$.

\begin{figure}[hp]
\centering
\includegraphics[width=0.55\textwidth]{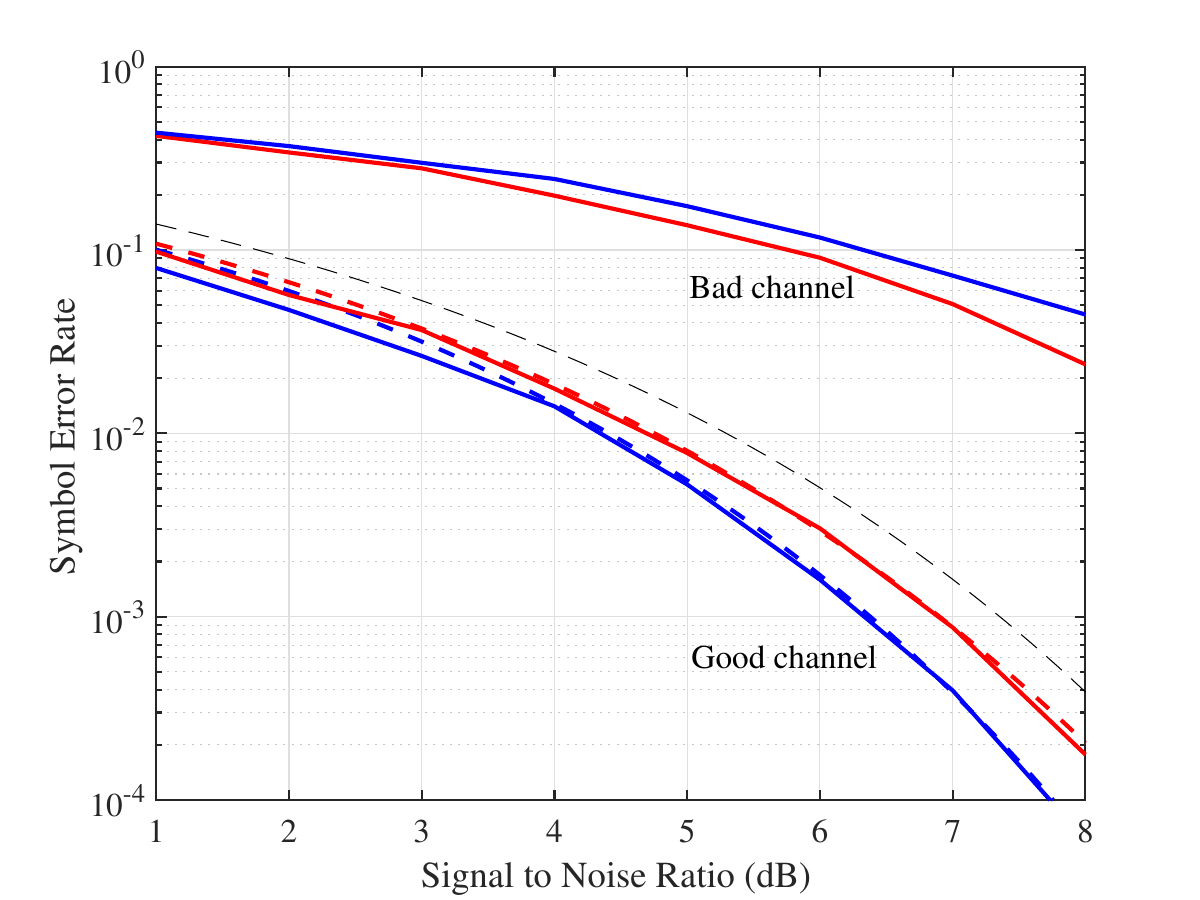}%
\caption{The distance spectrum upper bounds (dashed) and the simulation results (solid). Blue is the equidistant transform for the new signal set. Red is the transform $L4$ for 4-PSK signal set. Black is the standard transform for 4-PSK signal set.}
\label{fig_sig4}
\end{figure}

The distance spectrum upper bounds and simulation results are depicted in Fig.~\ref{fig_sig4}. Unfortunately, the synthetic bad channel is worse for this case due to its decreased minimum distance.

As a result of this section, we show that the synthetic good channel reaches the best possible distance spectrum upper bound with the new signal set for $q=4$ in 2-dimensions. It can be easily seen that $W^{++\dots+}$ channel is also optimal. As a consequence, for any $N$ and $K = 1$ the best non-binary ($q = 4$) polar code is defined in this way for two-dimensional signalling.
On the other hand the polar coding we have proposed for $4$-PSK in the previous section may provide superior error correction performance for $K> 2$ because it has a better synthetic bad channel than the code we proposed here with the new signal set. 
In this case, proposed method with the new signal set can be used for one-step polarization as a technique with $N=2$ and Rate=0.5 that can be seen as a new kind of repetition code for a signaling in two dimension.

A similar way can be followed to design signal-sets in 1-dimension for nonbinary equidistance. 
It is possible to consider a PAM-type new signal-set for $q=3$ in 1-dimension for nonbinary equidistance.

\begin{figure}[hp]
\centering
\vspace{0.01cm}
\begin{tikzpicture}[thick,scale=0.45*1.1*2.5*1.1*3.0*0.35, every node/.style={scale=1.1*1.4*1.25*0.5}]
\draw  (-2,0)  -- (2,0) ;
\draw  (0,-0.2)  -- (0,0.2) ;
\begin{scope}[very thin]
\end{scope}
\begin{scope}[very thin,dashed]
\end{scope}
\draw[fill=gray!160!white] (-1.866,0) circle (0.05cm) node[below ] {$s_0$} node[above] {$-1-\frac{\sqrt{3}}{2}$};
\begin{scope}[very thin,dashed]
\end{scope}
\draw[fill=gray!160!white] (-0.866,0.0) circle (0.05cm) node[below] {$s_1$} node[above] {$-\frac{\sqrt{3}}{2}$};
\draw[fill=gray!160!white] (1.866,0.0) circle (0.05cm) node[below] {$s_2$} node[above] {$1+\frac{\sqrt{3}}{2}$};
\end{tikzpicture}
\vspace{0.01cm}
\caption{The new signal-set for $q=3$ with $E_s=2.57$ joule/1-dimension for $q=3$ non-binary equidistance.}
\label{Fig_pam}
\end{figure}
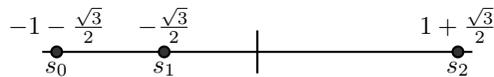

The new signal-set is obtained by the changing the geometry of the 3-PAM signaling as shown in Fig.~\ref{Fig_pam}. 
The following relation is found between the new Euclidean distances, depending on the amount of shift as a result of the rotation process. 
$$\|s_0-s_1\|=\alpha,\|s_1-s_2\|=\beta.$$
The following equation must be satisfied in order to have the equidistant characteristic of the transform given above. 
$$\sqrt{\|s_0-s_1\|^2+\|s_0-s_2\|^2}=\sqrt{\|s_1-s_2\|^2+\|s_1-s_2\|^2}$$ 
For this, $\beta=(1+\sqrt{3})\alpha$ is found and the amount of shift is determined.
As a result of these, the new signal-set is depicted in Fig.~\ref{Fig_pam}.

The distance properties are optimized by the nonbinary equidistance for the new signal-set as follows
The minimum distance is $d_{min}=2.415\sqrt{E_s}$
The distance spectrum is $N(d_{min})=2$.

This result provides the best distance spectrum bound for 1-dimensional signaling for $q=3$.

\section{Speed of Polarization}\label{S_speed_polarization}
\begin{figure}[hp]
\centering
\includegraphics[width=0.6\textwidth]{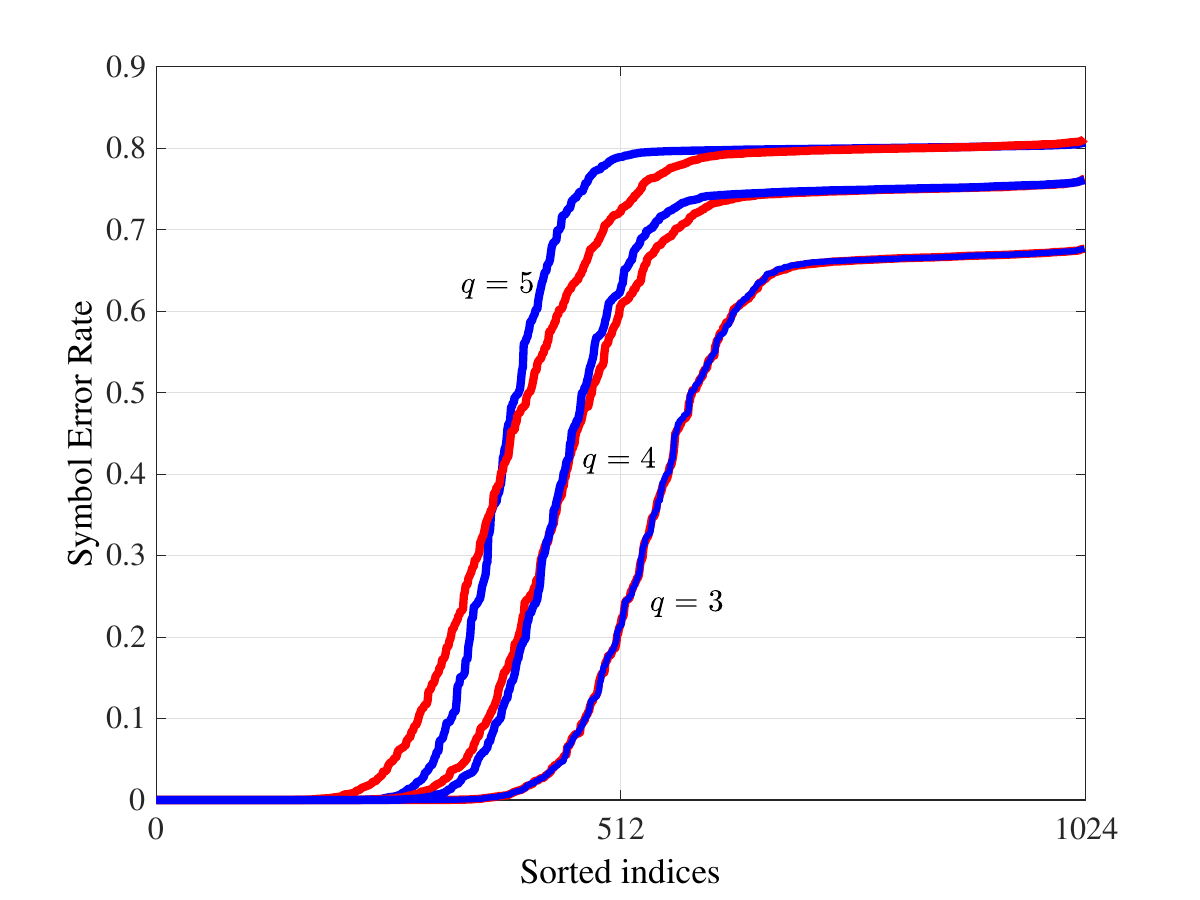}
\caption{Reliabilities of the indices of q-ary polar codes with blocklength $N=1024$ for q-ary PSK signal sets. Blue is for proposed transforms. Red is for the standard transforms.}
\label{Fig_rel}
\end{figure}

By the help of Monte-Carlo simulation, we investigate the speed of polarization for non-binary polar codes under a symbol based $q$-ary successive cancellation decoder. Reliabilities of the sorted synthetic channels are depicted in Fig.~\ref{Fig_rel}.
\newline As we can see from the figure, better distance properties lead to an improvement on the speed of polarization for non-binary polar codes.

\subsection{Experiments on the almost-equidistant transform}\label{S_speed_polarization_exp}
\begin{figure}[hp]
\centering
\begin{tikzpicture}[thick,scale=1.10*0.3*1.0, every node/.style={scale=1.10*0.45*2*1.0}]
\draw[->]  (-1,0)  -- (-0.2,0) ;
\draw  (0.7,0)  -- (1+1-0.5,0) ;
\draw[->]  (0.25,-2)  -- (0.25,-1.2) ;
\draw  (-1,-2)  -- (1+1-0.5,-2) ;
\draw (-0.2,-1.2)  -- (0.7,-1.2);
\draw (-0.2,0.5)  -- (0.7,0.5);
\draw (-0.2,-1.2)  -- (-0.2,0.5);
\draw (0.7,-1.2)  -- (0.7,0.5);
\begin{scope}[thick]
\draw (0.25,-0.35) node[] {$+$};
\end{scope}
\draw (-1,0) node[left] {$u_1$};
\draw (-1,-2) node[left] {$u_2$};
\draw (1+2-1.5,0)  -- (-1+4.5,0);
\draw (1+2-1.5,-2)  -- (-1+4.5,0-4);
\draw[->]  (-1,0-4)  -- (-0.2,0-4) ;
\draw  (0.7,0-4)  -- (1+1-0.5,0-4) ;
\draw[->]  (0.25,-2-4)  -- (0.25,-1.2-4) ;
\draw  (-1,-2-4)  -- (1+1-0.5,-2-4) ;
\draw (-0.2,-1.2-4)  -- (0.7,-1.2-4);
\draw (-0.2,0.5-4)  -- (0.7,0.5-4);
\draw (-0.2,-1.2-4)  -- (-0.2,0.5-4);
\draw (0.7,-1.2-4)  -- (0.7,0.5-4);
\begin{scope}[thick]
\draw (0.25,-0.35-4) node[] {$+$};
\end{scope}
\draw (-1,0-4) node[left] {$u_3$};
\draw (-1,-2-4) node[left] {$u_4$};
\draw (1+2-1.5,0-4)   -- (-1+4.5,-2);
\draw (1+2-1.5,-2-4)  -- (-1+4.5,-2-4);
\draw[->]  (-1+4.5,0)  -- (-0.2+4.5,0) ;
\draw  (0.7+4.5,0)  -- (1+1-0.5+4.5,0) ;
\draw[->]  (0.25+4.5,-2)  -- (0.25+4.5,-1.2) ;
\draw  (-1+4.5,-2)  -- (1+1-0.5+4.5,-2) ;
\draw (-0.2+4.5,-1.2)  -- (0.7+4.5,-1.2);
\draw (-0.2+4.5,0.5)  -- (0.7+4.5,0.5);
\draw (-0.2+4.5,-1.2)  -- (-0.2+4.5,0.5);
\draw (0.7+4.5,-1.2)  -- (0.7+4.5,0.5);
\begin{scope}[thick]
\draw (0.25+4.5,-0.35) node[] {$+$};
\end{scope}
%
\draw[->]  (-1+4.5,0-4)  -- (-0.2+4.5,0-4) ;
\draw  (0.7+4.5,0-4)  -- (1+1-0.5+4.5,0-4) ;
\draw[->]  (0.25+4.5,-2-4)  -- (0.25+4.5,-1.2-4) ;
\draw  (-1+4.5,-2-4)  -- (1+1-0.5+4.5,-2-4) ;
\draw (-0.2+4.5,-1.2-4)  -- (0.7+4.5,-1.2-4);
\draw (-0.2+4.5,0.5-4)  -- (0.7+4.5,0.5-4);
\draw (-0.2+4.5,-1.2-4)  -- (-0.2+4.5,0.5-4);
\draw (0.7+4.5,-1.2-4)  -- (0.7+4.5,0.5-4);
\begin{scope}[thick]
\draw (0.25+4.5,-0.35-4) node[] {$+$};
\end{scope}
\draw[->]  (-1,0-8)  -- (-0.2,0-8) ;
\draw  (0.7,0-8)  -- (1+1-0.5,0-8) ;
\draw[->]  (0.25,-2-8)  -- (0.25,-1.2-8) ;
\draw  (-1,-2-8)  -- (1+1-0.5,-2-8) ;
\draw (-0.2,-1.2-8)  -- (0.7,-1.2-8);
\draw (-0.2,0.5-8)  -- (0.7,0.5-8);
\draw (-0.2,-1.2-8)  -- (-0.2,0.5-8);
\draw (0.7,-1.2-8)  -- (0.7,0.5-8);
\begin{scope}[thick]
\draw (0.25,-0.35-8) node[] {$+$};
\end{scope}
\draw (-1,0-8) node[left] {$u_5$};
\draw (-1,-2-8) node[left] {$u_6$};
\draw (1+2-1.5,0-8)  -- (-1+4.5,0-8);
\draw (1+2-1.5,-2-8)  -- (-1+4.5,0-4-8);
\draw[->]  (-1,0-4-8)  -- (-0.2,0-4-8) ;
\draw  (0.7,0-4-8)  -- (1+1-0.5,0-4-8) ;
\draw[->]  (0.25,-2-4-8)  -- (0.25,-1.2-4-8) ;
\draw  (-1,-2-4-8)  -- (1+1-0.5,-2-4-8) ;
\draw (-0.2,-1.2-4-8)  -- (0.7,-1.2-4-8);
\draw (-0.2,0.5-4-8)  -- (0.7,0.5-4-8);
\draw (-0.2,-1.2-4-8)  -- (-0.2,0.5-4-8);
\draw (0.7,-1.2-4-8)  -- (0.7,0.5-4-8);
\begin{scope}[thick]
\draw (0.25,-0.35-4-8) node[] {$+$};
\end{scope}
\draw (-1,0-4-8) node[left] {$u_7$};
\draw (-1,-2-4-8) node[left] {$u_8$};
\draw (1+2-1.5,0-4-8)   -- (-1+4.5,-2-8);
\draw (1+2-1.5,-2-4-8)  -- (-1+4.5,-2-4-8);
\draw[->]  (-1+4.5,0-8)  -- (-0.2+4.5,0-8) ;
\draw  (0.7+4.5,0-8)  -- (1+1-0.5+4.5,0-8) ;
\draw[->]  (0.25+4.5,-2-8)  -- (0.25+4.5,-1.2-8) ;
\draw  (-1+4.5,-2-8)  -- (1+1-0.5+4.5,-2-8) ;
\draw (-0.2+4.5,-1.2-8)  -- (0.7+4.5,-1.2-8);
\draw (-0.2+4.5,0.5-8)  -- (0.7+4.5,0.5-8);
\draw (-0.2+4.5,-1.2-8)  -- (-0.2+4.5,0.5-8);
\draw (0.7+4.5,-1.2-8)  -- (0.7+4.5,0.5-8);
\begin{scope}[thick]
\draw (0.25+4.5,-0.35-8) node[] {$+$};
\end{scope}
%
\draw[->]  (-1+4.5,0-4-8)  -- (-0.2+4.5,0-4-8) ;
\draw  (0.7+4.5,0-4-8)  -- (1+1-0.5+4.5,0-4-8) ;
\draw[->]  (0.25+4.5,-2-4-8)  -- (0.25+4.5,-1.2-4-8) ;
\draw  (-1+4.5,-2-4-8)  -- (1+1-0.5+4.5,-2-4-8) ;
\draw (-0.2+4.5,-1.2-4-8)  -- (0.7+4.5,-1.2-4-8);
\draw (-0.2+4.5,0.5-4-8)  -- (0.7+4.5,0.5-4-8);
\draw (-0.2+4.5,-1.2-4-8)  -- (-0.2+4.5,0.5-4-8);
\draw (0.7+4.5,-1.2-4-8)  -- (0.7+4.5,0.5-4-8);
\begin{scope}[thick]
\draw (0.25+4.5,-0.35-4-8) node[] {$+$};
\end{scope}
\draw (1+2-1.5+9/2,0)  -- (-1+4.5+9/2,0);
\draw (1+2-1.5+9/2,-2)  -- (-1+4.5+9/2,0-4);

\draw (1+2-1.5+9/2,0-4)   -- (-1+4.5+9/2,-8);
\draw (1+2-1.5+9/2,-2-4)  -- (-1+4.5+9/2,-2-4-6);
\draw[->]  (-1+4.5+9/2,0)  -- (-0.2+4.5+9/2,0) ;
\draw[->]  (0.7+4.5+9/2,0)  -- (1+1-0.5+4.5+9/2,0) ;
\draw[->]  (0.25+4.5+9/2,-2)  -- (0.25+4.5+9/2,-1.2) ;
\draw[->]  (-1+4.5+9/2,-2)  -- (1+1-0.5+4.5+9/2,-2) ;
\draw (-0.2+4.5+9/2,-1.2)  -- (0.7+4.5+9/2,-1.2);
\draw (-0.2+4.5+9/2,0.5)  -- (0.7+4.5+9/2,0.5);
\draw (-0.2+4.5+9/2,-1.2)  -- (-0.2+4.5+9/2,0.5);
\draw (0.7+4.5+9/2,-1.2)  -- (0.7+4.5+9/2,0.5);
\begin{scope}[thick]
\draw (0.25+4.5+9/2,-0.35) node[] {$f$};
\end{scope}
\draw (1+2-1.5+4.5+9/2,0-0.03) node[right] {$x_1$};
\draw (1+2-1.5+4.5+9/2,-2-0.03) node[right] {$x_2$};
\draw[->]  (-1+4.5+9/2,0-4)  -- (-0.2+4.5+9/2,0-4) ;
\draw[->]  (0.7+4.5+9/2,0-4)  -- (1+1-0.5+4.5+9/2,0-4) ;
\draw[->]  (0.25+4.5+9/2,-2-4)  -- (0.25+4.5+9/2,-1.2-4) ;
\draw[->]  (-1+4.5+9/2,-2-4)  -- (1+1-0.5+4.5+9/2,-2-4) ;
\draw (-0.2+4.5+9/2,-1.2-4)  -- (0.7+4.5+9/2,-1.2-4);
\draw (-0.2+4.5+9/2,0.5-4)  -- (0.7+4.5+9/2,0.5-4);
\draw (-0.2+4.5+9/2,-1.2-4)  -- (-0.2+4.5+9/2,0.5-4);
\draw (0.7+4.5+9/2,-1.2-4)  -- (0.7+4.5+9/2,0.5-4);
\begin{scope}[thick]
\draw (0.25+4.5+9/2,-0.35-4) node[] {$f$};
\end{scope}
\draw (1+2-1.5+4.5+9/2,0-0.03-4) node[right] {$x_3$};
\draw (1+2-1.5+4.5+9/2,-2-0.03-4) node[right] {$x_4$};
\draw (1+2-1.5+9/2,0-8)  -- (-1+4.5+9/2,0-8+6);
\draw (1+2-1.5+9/2,-2-8)  -- (-1+4.5+9/2,0-4-8+6);
\draw (1+2-1.5+9/2,0-4-8)   -- (-1+4.5+9/2,-2-8);
\draw (1+2-1.5+9/2,-2-4-8)  -- (-1+4.5+9/2,-2-4-8);
\draw[->]  (-1+4.5+9/2,0-8)  -- (-0.2+4.5+9/2,0-8) ;
\draw[->]  (0.7+4.5+9/2,0-8)  -- (1+1-0.5+4.5+9/2,0-8) ;
\draw[->]  (0.25+4.5+9/2,-2-8)  -- (0.25+4.5+9/2,-1.2-8) ;
\draw[->]  (-1+4.5+9/2,-2-8)  -- (1+1-0.5+4.5+9/2,-2-8) ;
\draw (-0.2+4.5+9/2,-1.2-8)  -- (0.7+4.5+9/2,-1.2-8);
\draw (-0.2+4.5+9/2,0.5-8)  -- (0.7+4.5+9/2,0.5-8);
\draw (-0.2+4.5+9/2,-1.2-8)  -- (-0.2+4.5+9/2,0.5-8);
\draw (0.7+4.5+9/2,-1.2-8)  -- (0.7+4.5+9/2,0.5-8);
\begin{scope}[thick]
\draw (0.25+4.5+9/2,-0.35-8) node[] {$f$};
\end{scope}
\draw (1+2-1.5+4.5+9/2,0-0.03-8) node[right] {$x_5$};
\draw (1+2-1.5+4.5+9/2,-2-0.03-8) node[right] {$x_6$};
\draw[->]  (-1+4.5+9/2,0-4-8)  -- (-0.2+4.5+9/2,0-4-8) ;
\draw[->]  (0.7+4.5+9/2,0-4-8)  -- (1+1-0.5+4.5+9/2,0-4-8) ;
\draw[->]  (0.25+4.5+9/2,-2-4-8)  -- (0.25+4.5+9/2,-1.2-4-8) ;
\draw[->]  (-1+4.5+9/2,-2-4-8)  -- (1+1-0.5+4.5+9/2,-2-4-8) ;
\draw (-0.2+4.5+9/2,-1.2-4-8)  -- (0.7+4.5+9/2,-1.2-4-8);
\draw (-0.2+4.5+9/2,0.5-4-8)  -- (0.7+4.5+9/2,0.5-4-8);
\draw (-0.2+4.5+9/2,-1.2-4-8)  -- (-0.2+4.5+9/2,0.5-4-8);
\draw (0.7+4.5+9/2,-1.2-4-8)  -- (0.7+4.5+9/2,0.5-4-8);
\begin{scope}[thick]
\draw (0.25+4.5+9/2,-0.35-4-8) node[] {$f$};
\end{scope}
\draw (1+2-1.5+4.5+9/2,0-0.03-4-8) node[right] {$x_7$};
\draw (1+2-1.5+4.5+9/2,-2-0.03-4-8) node[right] {$x_8$};
\end{tikzpicture}
\caption{An encoder scheme of non-binary polar codes with $f$ at only the channel stage. Other stages have the standard transform.}
\label{Fig_chst}
\end{figure}
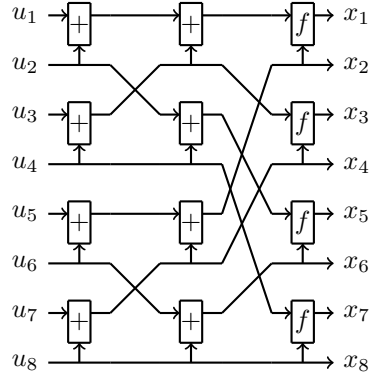

We experimentally investigate the contribution of equidistant polarization transforms that are placed at only the channel stage as shown in Fig.~\ref{Fig_chst}. 

\begin{figure}[hp]
\centering
\includegraphics[width=0.6\textwidth]{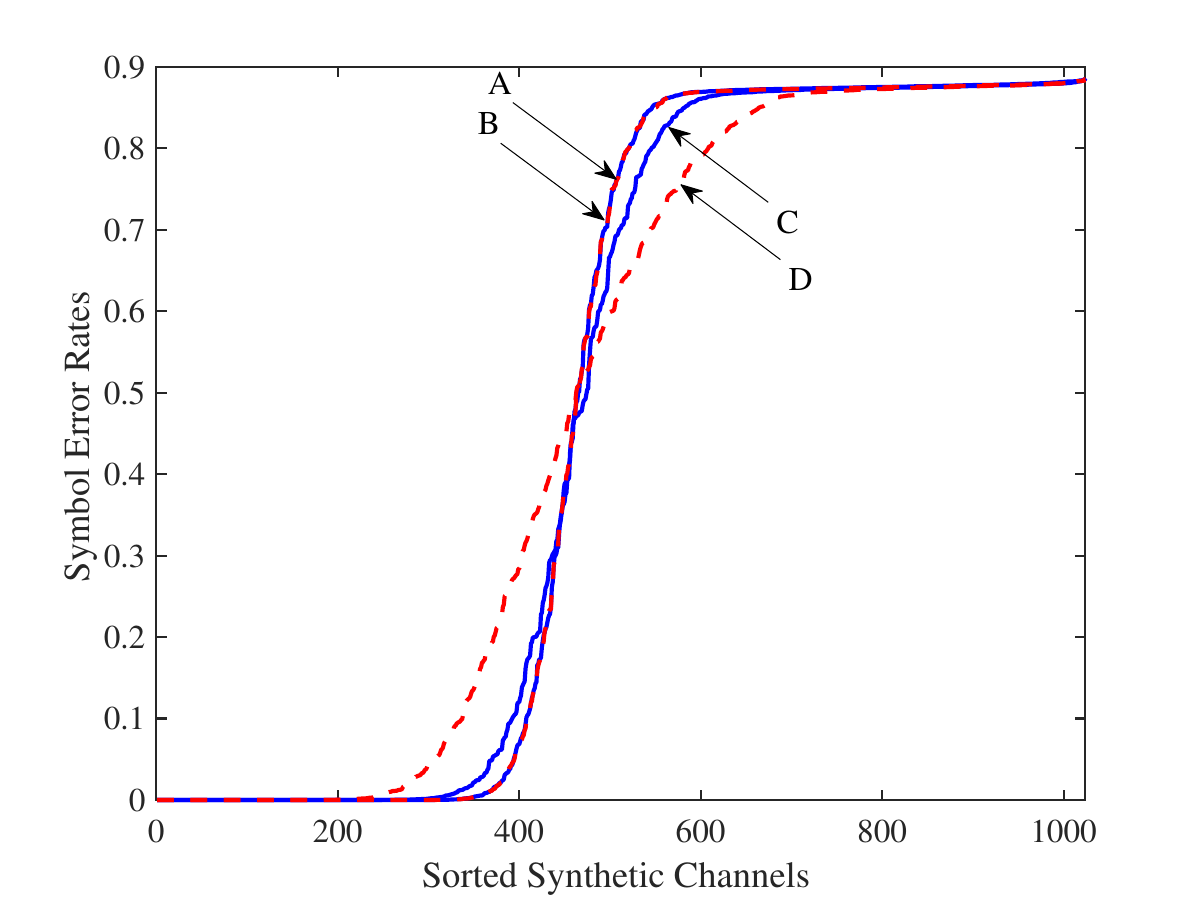}
\caption{Reliabilities of indices for $N=1024$ 8-ary polar codes. A-all stages with $L_8$, B-channel stage with $L_8$, others with standard,C-all stages with (\ref{polaproc_1}), D-channel stage with (\ref{polaproc_1}), others with standard. }
\label{fig_fin}
\end{figure}

It was depicted in Fig.~\ref{fig_fin} that contribution of any polarization transforms from the first stage to the channel state are insignificant when the equidistant transform is placed at the channel stage. This result is not true for other (not equidistant) polarization transforms. 

The equidistant channels are first discussed for polar coding in the related work \cite{sasoglu_polarization_2009}. It was mentioned that the standard transform polarizes equidistant channels, regardless of the input alphabet size. 
By similar argument, the equidistant transform that is placed at the channel stage creates an equidistant synthetic good channel, and hence, the standard transforms that are placed at all of the previous stages provide the polarization, regardless of the input alphabet size. 
This assertion is provided by the computer simulation in Fig.~\ref{fig_fin}.     

\pagebreak

\section{Asymptotic Behaviour of the Equidistant Property}\label{S_asymptotic}
\begin{figure}[hp]
\centering
\includegraphics[width=0.4\textwidth]{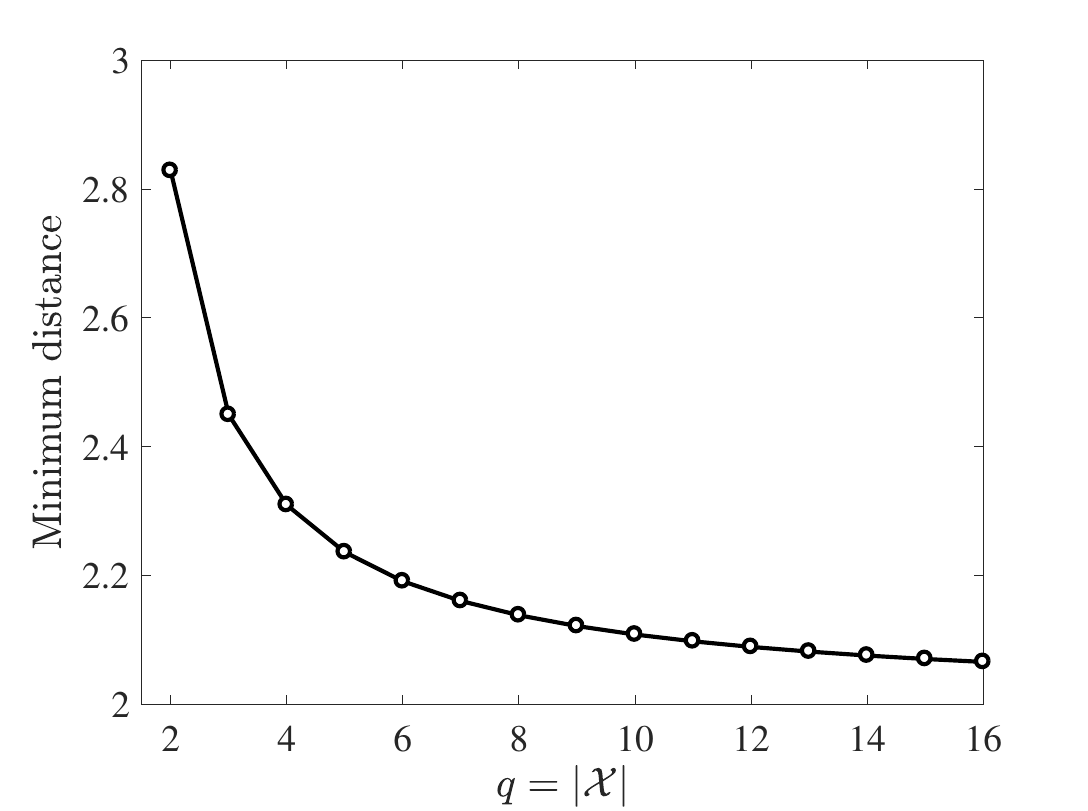}
\caption{The minimum distances of the equidistant transforms for $q$-ary PSK signal set with $E_s=1$ Joule/2-dimensions.} 
\label{Fig_dmin}
\end{figure}

The minimum distances of the equidistant transforms are plotted for $q$-ary PSK signal set in Fig.~\ref{Fig_dmin}.  
\newline The minimum distance of the standard transform and the transform in (\ref{polaproc_1}) is 
$$\lim_{q\rightarrow \infty} d_{min}=0$$ 
for $q$-ary PSK signal set.
Notice that the minimum distance of the equidistant transforms is $$\lim_{q\rightarrow \infty} d_{min}=2\sqrt{E_s}$$ for $q$-ary PSK signal set.

Furthermore, the equidistant distant spectrum upper bound for $q\rightarrow \infty$ is 
$$P_e\leq (q-1)Q\left(\sqrt{2SNR}\right) $$ 
for $q$-ary PSK signal set.

The asymptotic behavior of the equidistant property provides that the non-binary equidistant polar coding is a promising way for very high order input alphabet sizes.
It can be seen that the ultimate limit of the minimum distance $\lim_{q\rightarrow \infty} d_{min}=2\sqrt{E_s}$ can be approached for high order of $q$ and $q$-ary PSK signal set by the polarizing transform $f(u_1,u_2)=u_1\oplus\pi(u_2)$. For this case, we provide the following toe example for 10-PSK: 

\begin{myexample}
For 10-PSK signal set, a table can be constructed as follows. The effective minimum distance of this example is close to the $d=2\sqrt{E_s}$ which is ultimate limit for $q\rightarrow \infty$. Notice that this is close to the optimal $d_{min}$ for $q=10$ and 10-PSK signal set. 
\begin{center}
\begin{tikzpicture}[scale=0.95, every node/.style={scale=0.95}]
\draw(-3.5,-2.00) node[] {$L10:\pi=\left(\begin{smallmatrix} 0& 1& 2& 3& 4& 5& 6& 7& 8& 9\\ 0&4&7&2&9&5&1&8&3&6 \end{smallmatrix}\right)$};\hspace{1cm}
\draw  (-1.0, 0.0)  -- (4, 0.0) ;
\draw  (-1.0,-0.5)  -- (4,-0.5) ;
\draw  (-1.0,-1.0)  -- (4,-1.0) ;
\draw  (-1.0,-1.5)  -- (4,-1.5) ;
\draw  (-1.0,-2.0)  -- (4,-2.0) ;
\draw  (-1.0,-2.5)  -- (4,-2.5) ;
\draw  (-1.0,-3.0)  -- (4,-3.0) ;
\draw  (-1.0,-3.5)  -- (4,-3.5) ;
\draw  (-1.0,-4.0)  -- (4,-4.0) ;
\draw  (-1.0,-4.5)  -- (4,-4.5) ;
\draw  (-1.0,-5.0)  -- (4,-5.0) ;
\draw  (-1.0, 0.0)  -- (-1.0,-5) ;
\draw  (-0.5, 0.0)  -- (-0.5,-5) ;
\draw  (-0.0, 0.0)  -- (-0.0,-5) ;
\draw  ( 0.5, 0.0)  -- ( 0.5,-5) ;
\draw  ( 1.0, 0.0)  -- ( 1.0,-5) ;
\draw  ( 1.5, 0.0)  -- ( 1.5,-5) ;
\draw  ( 2.0, 0.0)  -- ( 2.0,-5) ;
\draw  ( 2.5, 0.0)  -- ( 2.5,-5) ;
\draw  ( 3.0, 0.0)  -- ( 3.0,-5) ;
\draw  ( 3.5, 0.0)  -- ( 3.5,-5) ;
\draw  ( 4.0, 0.0)  -- ( 4.0,-5) ;
\draw[fill=gray!50] (-1.25+0.5-0.25,-0.5) rectangle (-1.25+0.5-0.25+0.5,-0.5+0.5) (-1.25+0.5+0.25-0.25,-0.5) node[above]  {$00$};%
\draw (-1.25+0.5,-1.0) node[above] {$61$};
\draw (-1.25+0.5,-1.5) node[above] {$32$};
\draw (-1.25+0.5,-2.0) node[above] {$83$};
\draw (-1.25+0.5,-2.5) node[above] {$14$};
\draw (-1.25+0.5,-3.0) node[above] {$55$};
\draw (-1.25+0.5,-3.5) node[above] {$96$};
\draw (-1.25+0.5,-4.0) node[above] {$27$};
\draw (-1.25+0.5,-4.5) node[above] {$78$};
\draw (-1.25+0.5,-5.0) node[above] {$49$};%
\draw (-1.25+1.0,-0.5) node[above] {$10$};
\draw (-1.25+1.0,-1.0) node[above] {$71$};
\draw (-1.25+1.0,-1.5) node[above] {$42$};
\draw[fill=gray!0] (-1.25+1.0-0.25,-2.0) rectangle (-1.25+1.0-0.25+0.5,-2.0+0.5) (-1.25+1.0+0.25-0.25,-2.0) node[above]  {$93$};%
\draw (-1.25+1.0,-2.5) node[above]  {$24$};
\draw (-1.25+1.0,-3.0) node[above]  {$65$};
\draw[fill=gray!50] (-1.25+1.0-0.25,-3.5) rectangle (-1.25+1.0-0.25+0.5,-3.5+0.5) (-1.25+1.0-0.25+0.25,-3.5) node[above]  {$06$};
\draw (-1.25+1.0,-4.0) node[above]  {$37$};
\draw (-1.25+1.0,-4.5) node[above]  {$88$};
\draw (-1.25+1.0,-5.0) node[above]  {$59$};%
\draw (-1.25+1.5,-0.5) node[above] {$20$};
\draw (-1.25+1.5,-1.0) node[above] {$81$};
\draw (-1.25+1.5,-1.5) node[above] {$52$};
\draw[fill=gray!50] (-1.25+1.5-0.25,-2.0) rectangle (-1.25+1.5-0.25+0.5,-2.0+0.5) (-1.25+1.5-0.25+0.25,-2.0) node[above] {$03$};
\draw (-1.25+1.5,-2.5) node[above] {$34$};
\draw (-1.25+1.5,-3.0) node[above] {$75$};
\draw[fill=gray!0] (-1.25+1.5-0.25,-3.5) rectangle (-1.25+1.5-0.25+0.5,-3.5+0.5) (-1.25+1.5+0.25-0.25,-3.5) node[above] {$16$};%
\draw (-1.25+1.5,-4.0) node[above] {$47$};
\draw (-1.25+1.5,-4.5) node[above] {$98$};
\draw (-1.25+1.5,-5.0) node[above] {$69$};%
\draw (-1.25+2,-0.5) node[above]    {$30$};
\draw[fill=gray!0] (-1.25+2-0.25,-1.0) rectangle (-1.25+2-0.25+0.5,-1.0+0.5) (-1.25+2+0.25-0.25,-1.0) node[above]     {$91$};%
\draw (-1.25+2,-1.5) node[above]    {$62$};
\draw (-1.25+2,-2.0) node[above]    {$13$};
\draw (-1.25+2,-2.5) node[above]    {$44$};
\draw (-1.25+2,-3.0) node[above]    {$85$};
\draw (-1.25+2,-3.5) node[above]    {$26$};
\draw (-1.25+2,-4.0) node[above]    {$57$};
\draw[fill=gray!50] (-1.25+2-0.25,-4.5) rectangle (-1.25+2-0.25+0.5,-4.5+0.5) (-1.25+2-0.25+0.25,-4.5) node[above]    {$08$};
\draw (-1.25+2,-5.0) node[above]    {$79$};%
\draw (-1.25+2.5,-0.5) node[above] {$40$};
\draw[fill=gray!50] (-1.25+2.5-0.25,-1.0) rectangle (-1.25+2.5-0.25+0.5,-1.0+0.5) (-1.25+2.5-0.25+0.25,-1.0) node[above] {$01$};
\draw (-1.25+2.5,-1.5) node[above] {$72$};
\draw (-1.25+2.5,-2.0) node[above] {$23$};
\draw[fill=gray!0] (-1.25+2.5-0.25,-2.5) rectangle (-1.25+2.5-0.25+0.5,-2.5+0.5) (-1.25+2.5+0.25-0.25,-2.5) node[above] {$54$};%
\draw (-1.25+2.5,-3.0) node[above] {$95$};
\draw (-1.25+2.5,-3.5) node[above] {$36$};
\draw (-1.25+2.5,-4.0) node[above] {$67$};
\draw (-1.25+2.5,-4.5) node[above] {$18$};
\draw (-1.25+2.5,-5.0) node[above] {$89$};%
\draw (-1.25+3,-0.5) node[above]    {$50$};
\draw (-1.25+3,-1.0) node[above]    {$11$};
\draw (-1.25+3,-1.5) node[above]    {$82$};
\draw (-1.25+3,-2.0) node[above]    {$33$};
\draw (-1.25+3,-2.5) node[above]    {$64$};
\draw[fill=gray!50] (-1.25+3-0.25,-3.0) rectangle (-1.25+3-0.25+0.5,-3.0+0.5) (-1.25+3-0.25+0.25,-3.0) node[above]    {$05$};
\draw (-1.25+3,-3.5) node[above]    {$46$};
\draw[fill=gray!0] (-1.25+3-0.25,-4.0) rectangle (-1.25+3-0.25+0.5,-4.0+0.5) (-1.25+3+0.25-0.25,-4.0) node[above]    {$77$};%
\draw (-1.25+3,-4.5) node[above]    {$28$};
\draw (-1.25+3,-5.0) node[above]    {$99$};%
\draw (-1.25+3.5,-0.5) node[above] {$60$};
\draw (-1.25+3.5,-1.0) node[above] {$21$};
\draw[fill=gray!0] (-1.25+3.5-0.25,-1.5) rectangle (-1.25+3.5-0.25+0.5,-1.5+0.5) (-1.25+3.5+0.25-0.25,-1.5) node[above] {$92$};%
\draw (-1.25+3.5,-2.0) node[above] {$43$};
\draw (-1.25+3.5,-2.5) node[above] {$74$};
\draw (-1.25+3.5,-3.0) node[above] {$15$};
\draw (-1.25+3.5,-3.5) node[above] {$56$};
\draw (-1.25+3.5,-4.0) node[above] {$87$};
\draw (-1.25+3.5,-4.5) node[above] {$38$};
\draw[fill=gray!50] (-1.25+3.5-0.25,-5.0) rectangle (-1.25+3.5-0.25+0.5,-5.0+0.5) (-1.25+3.5-0.25+0.25,-5.0) node[above] {$09$};%
\draw (-1.25+4,-0.5) node[above]    {$70$};
\draw (-1.25+4,-1.0) node[above]    {$31$};
\draw[fill=gray!50] (-1.25+4-0.25,-1.5) rectangle(-1.25+4-0.25+0.5,-1.5+0.5) (-1.25+4-0.25+0.25,-1.5)  node[above]    {$02$};
\draw (-1.25+4,-2.0) node[above]    {$53$};
\draw (-1.25+4,-2.5) node[above]    {$84$};
\draw[fill=gray!0] (-1.25+4-0.25,-3.0) rectangle (-1.25+4-0.25+0.5,-3.0+0.5) (-1.25+4+0.25-0.25,-3.0) node[above]    {$25$};%
\draw (-1.25+4,-3.5) node[above]    {$66$};
\draw (-1.25+4,-4.0) node[above]    {$97$};
\draw (-1.25+4,-4.5) node[above]    {$48$};
\draw (-1.25+4,-5.0) node[above]    {$19$};%
\draw (-1.25+4.5,-0.5) node[above]    {$80$};
\draw (-1.25+4.5,-1.0) node[above]    {$41$};
\draw (-1.25+4.5,-1.5) node[above]    {$12$};
\draw (-1.25+4.5,-2.0) node[above]    {$63$};
\draw (-1.25+4.5,-2.5) node[above]    {$94$};
\draw[fill=gray!0] (-1.25+4.5-0.25,-3.0) rectangle (-1.25+4.5-0.25+0.5,-3.0+0.5) (-1.25+4.5+0.25-0.25,-3.0) node[above]    {$35$};%
\draw (-1.25+4.5,-3.5) node[above]    {$76$};
\draw[fill=gray!50] (-1.25+4.5-0.25,-4.0) rectangle(-1.25+4.5-0.25+0.5,-4.0+0.5) (-1.25+4.5-0.25+0.25,-4.0)  node[above]    {$07$};
\draw (-1.25+4.5,-4.5) node[above]    {$58$};
\draw (-1.25+4.5,-5.0) node[above]    {$29$};%
\draw (-1.25+5,-0.5) node[above]    {$90$};
\draw (-1.25+5,-1.0) node[above]    {$51$};
\draw (-1.25+5,-1.5) node[above]    {$22$};
\draw (-1.25+5,-2.0) node[above]    {$73$};
\draw[fill=gray!50] (-1.25+5-0.25,-2.5) rectangle (-1.25+5-0.25+0.5,-2.5+0.5) (-1.25+5-0.25 + 0.25,-2.5) node[above]   {$04$};
\draw[fill=gray!0] (-1.25+5-0.25,-3.0) rectangle (-1.25+5-0.25+0.5,-3.0+0.5) (-1.25+5+0.25-0.25,-3.0) node[above]    {$45$};%
\draw (-1.25+5,-3.5) node[above]    {$86$};
\draw (-1.25+5,-4.0) node[above]    {$17$};
\draw (-1.25+5,-4.5) node[above]    {$68$};
\draw (-1.25+5,-5.0) node[above]    {$39$};
\end{tikzpicture}
\end{center}
\end{myexample}
Here, we can give a short comment on the existence of polarizing transforms with $d_{min}=2\sqrt{E_s}$ for high order of $q$. A geometric property can be noticed in Fig.~\ref{Fig_8psk}. that the number of pairs $(s_i,s_k)$ with the following equality increases for high order of $q$. 
$$\sqrt{\|s_0-s_i \|^2+\|s_0-s_k \|^2}=2\sqrt{E_s}$$
Moreover, there are some extra equalities in (\ref{sigse}) that can help to approach the ultimate limit for PSK signal set with high order of $q$.   
Intuitively, this result can provide that the proper transforms can be found for $q\rightarrow \infty$. This can be seen as a well known problem in lattice theory or coloring map problem in graph theory.  

In the next section, we provide some simulation results to investigate the error performance for $q$-ary PSK signal sets.

%
%
%

\vspace{0.5cm}
\newpage
\section{Simulation Results: Experiments to investigate the error performance}\label{S_simulation_result}
We are investigating the error performance of non-binary polar codes for AWGN channels under the $q$-ary successive cancellation decoder. These codes were constructed using Monte-Carlo simulations for $2$ dB SNR. Spectral efficiency is set to $1$ bit/channel-use.

\begin{figure}[hp]
\centering
\includegraphics[width=0.7\textwidth]{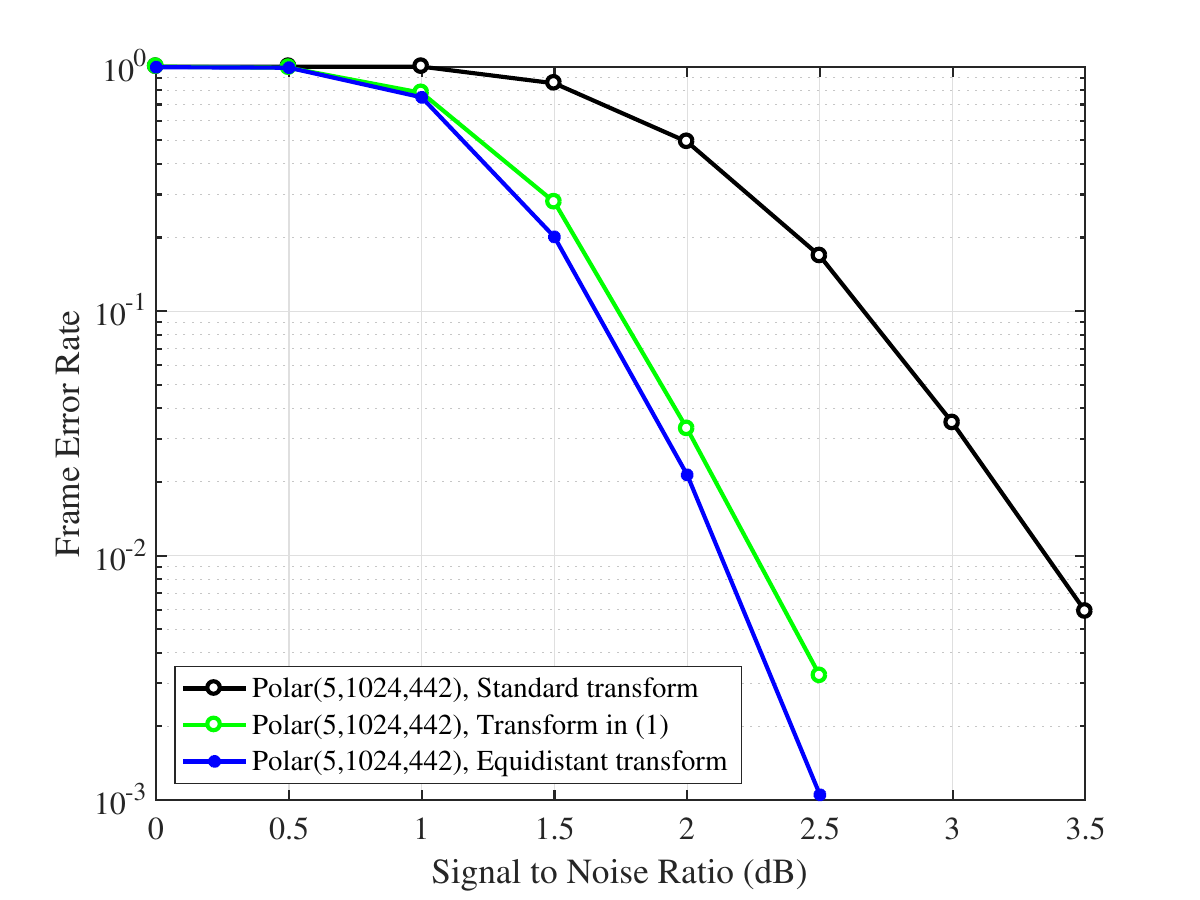}%
\caption{Frame error rates of non-binary polar codes for 5-PSK signal set.}
\label{fig_fer1}
\end{figure}

The frame error rates are depicted in Fig.\ref{fig_fer1} for $q=5$, and 5-PSK signalling. The equidistant transform outperforms the standard transform and the transform in (\ref{polaproc_1}). 

To investigate the error rates that are shown in Fig.\ref{fig_fer1}, we can consider the following distance profiles for the most common two error events in this case. 
\begin{itemize}
\item By using standard transform, there are two error events within the distance $d=1.663\sqrt{E_s}$.   
\item By using the equidistant transform, there are two error events within the distance $d=2.236\sqrt{E_s}$
\item Distance profile depends on the transmitted signal for the transform in (\ref{polaproc_1}). There is one signal with two error events within the distance $d=2.236\sqrt{E_s}$. Other four signals with one error event within the distance $d=1.663\sqrt{E_s}$ and one error event within the distance $d=2.236\sqrt{E_s}$. We suppose that the probability is equal to $1/q$ for each transmitted signal. There are two error events within the effective distance $d=\frac{(2.236+2.236)+4\times(1.663+2.236)}{5\times2}=2\sqrt{E_s}$.
\end{itemize} 
These results show that standard transform provides poor performances. Performance of Transform in (\ref{polaproc_1}) is closer to the performance of the equidistant transform.  

\begin{figure}[hp]
\centering
\includegraphics[width=0.7\textwidth]{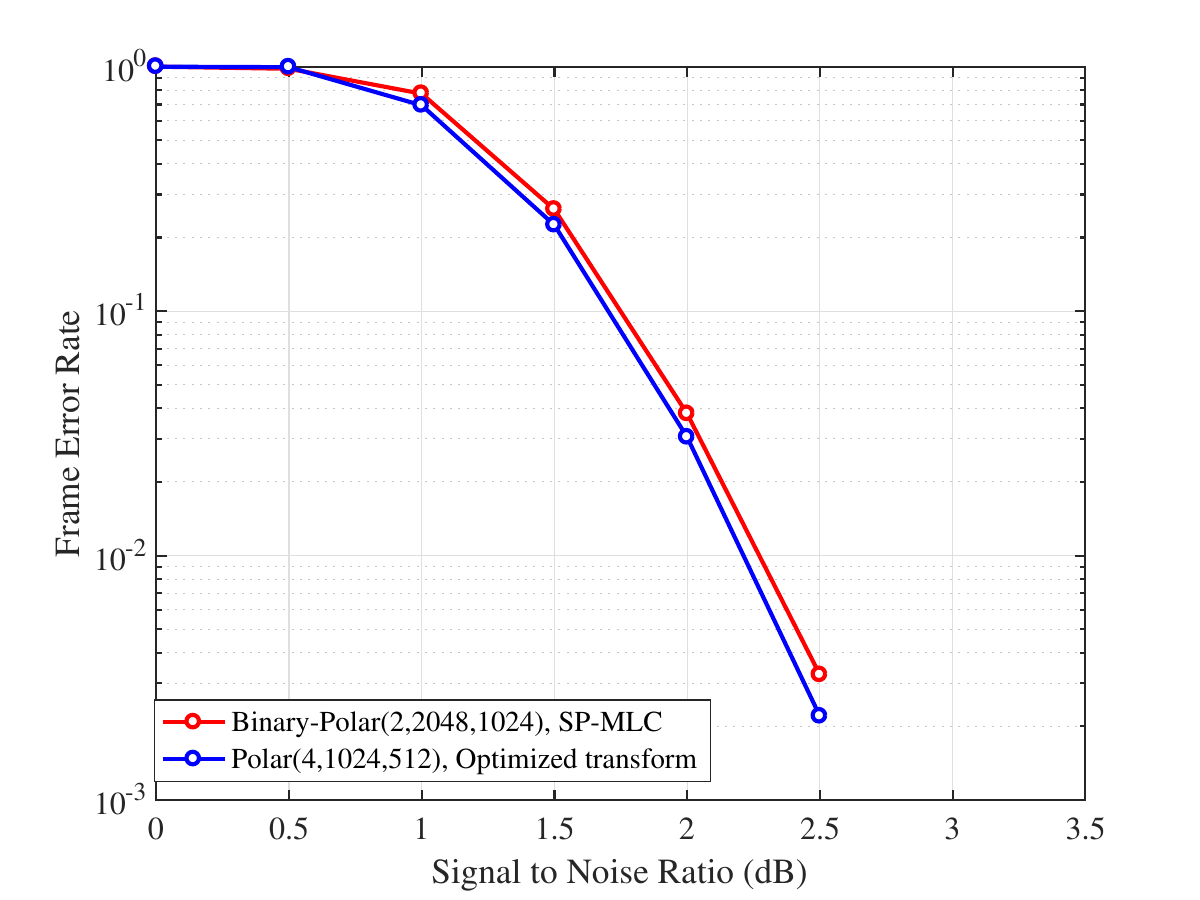}%
\caption{Frame error rates of non-binary polar code and SP-MLC binary polar code for 4-PSK signal set.}
\label{fig_fer2}
\end{figure}

In the next experiment, the frame error rates are provided in Fig.\ref{fig_fer2} for the optimized transform for $q=4$, and 4-PSK signal set. 
Note that we compare our results with the set partitioned (SP) binary polar codes \cite{seidl_polar-coded_2013} which are state-of-the-art transmission schemes with multi-level coding (MLC) for polar codes for 4-PSK signalling.
The optimized transform outperforms state-of-the-art MLC. 
\pagebreak

Furthermore, proposed non-binary codes in this study can also be adopted to the set partitioned MLC schemes to improve the error performance. This is out of the scope in this paper. We left this research direction to the future work. 

\begin{figure}[hp]
\centering
\includegraphics[width=0.7\textwidth]{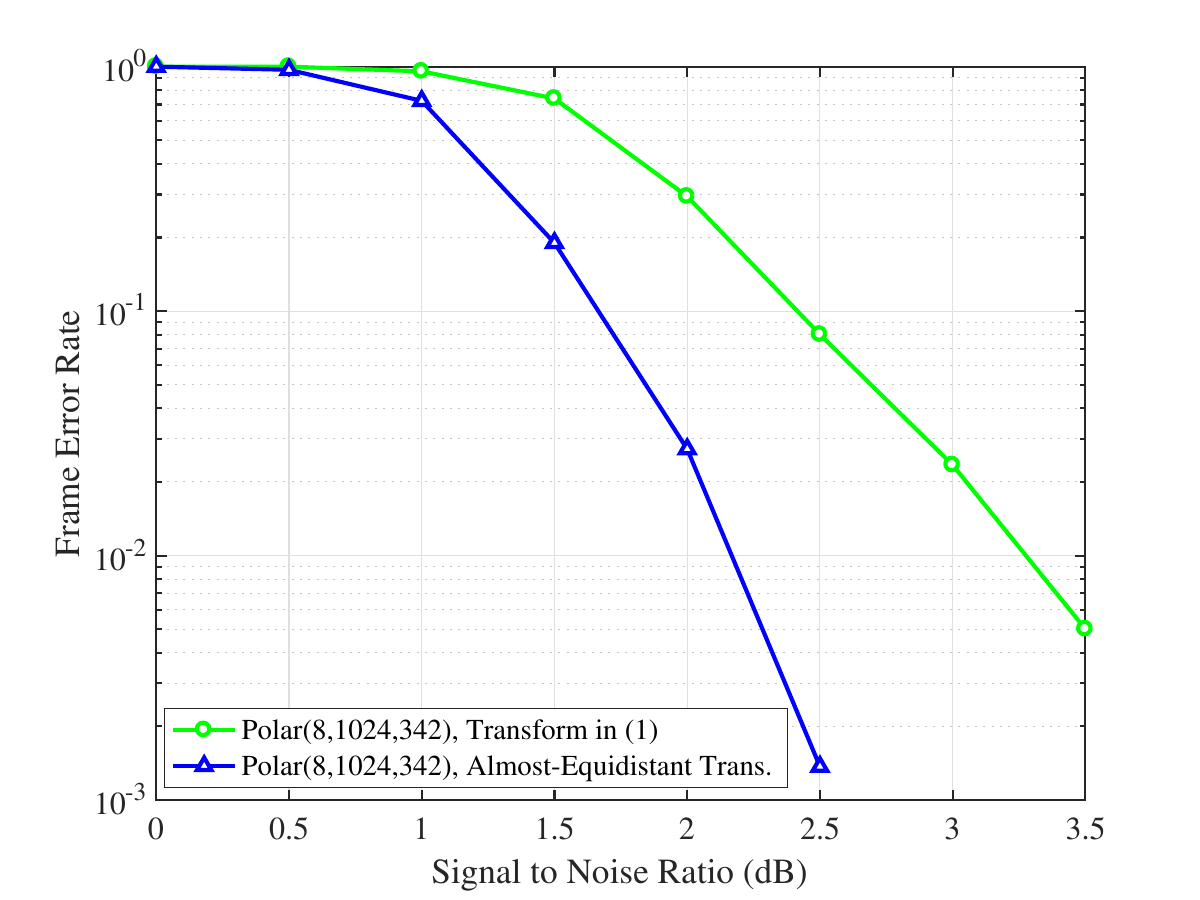}%
\caption{Frame error rates of non-binary polar codes for 8-PSK signal set.}
\label{fig_fer3}
\end{figure}

The frame error rates are depicted in Fig.\ref{fig_fer3} for $q=8$, and 8-PSK signalling. The almost-equidistant transform outperforms the transform in (\ref{polaproc_1}). 

To investigate the error rates that are shown in Fig.\ref{fig_fer3}, we can consider the following distance profiles for the most common error event in this case. 
\begin{itemize}
\item By using standard transform, there are two error events within the distance $d=1.082\sqrt{E_s}$.   
\item By using the almost-equidistant transform, there is error event within the distance $d=2\sqrt{E_s}$
\item Distance profile depends on the transmitted signal for the transform in (\ref{polaproc_1}). There is one signal with error event within the distance $d=2\sqrt{E_s}$. Other seven signals with error event within the distance $d=1.082\sqrt{E_s}$ and one error event within the distance $d=2\sqrt{E_s}$. We suppose that the probability is equal to $1/q$ for each transmitted signal. There are two error events within the effective distance $d=\frac{(2+7\times1.082)}{8}=1.197\sqrt{E_s}$.
\end{itemize} 
These results show that performance of the equidistant transform is significantly better than the performance of Transform in (\ref{polaproc_1}) which is closer to the standard transform for $q=8$.  

\begin{figure}[hp]
\centering
\includegraphics[width=0.7\textwidth]{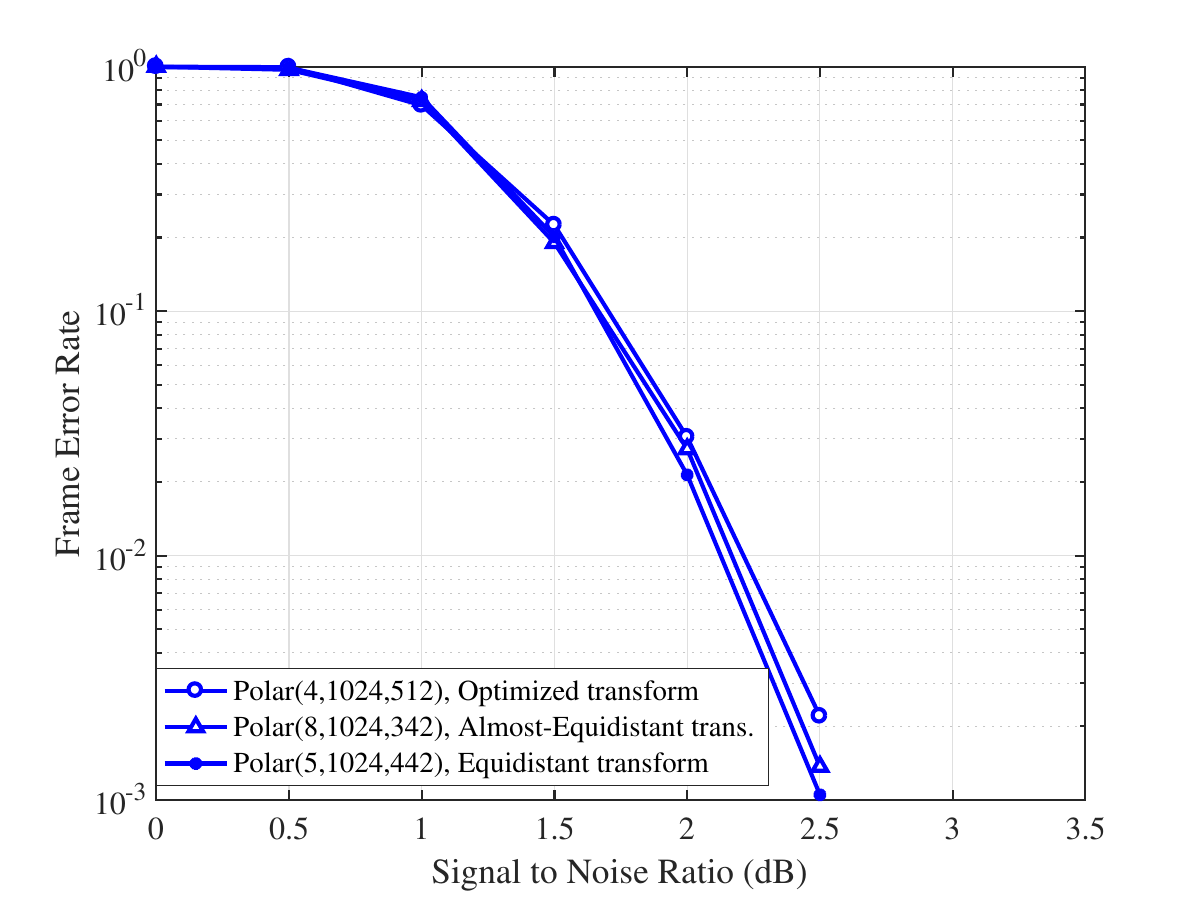}%
\caption{Frame error rates of the proposed non-binary polar codes for 4, 5 and 8-PSK signal sets.}
\label{fig_fer4}
\end{figure}

In the next experiment, the frame error rates are provided in Fig.\ref{fig_fer4} for the optimized transform for $q=4$, and 4-PSK signal set, the equidistant transform for $q=5$, and 5-PSK signal set, the almost-equidistant transform for $q=8$, and 8-PSK signal set.
The equidistant transform outperforms the optimized and almost equidistant transforms.

Results in Fig.\ref{fig_fer2} and Fig.\ref{fig_fer4} show that performance of SP-MLC scheme by Siedl can be obtained by non-binary polar coding for $q=4,5,8$ with careful design of polarizing transforms.

\section{Conclusion}\label{S_conclusion}
We conclude that PSK type signal set can provide equidistant property by using proper polarizing transforms. We investigate the connection between the channel polarization with non-binary transform and the signal set design for equidistant channels. 
We show that the polarization speed can be controlled by distance profile of polarizing transform.  
An asymptotic behaviour of the equidistant property show that minimum distance goes to $2\sqrt{E_s}$ for $q\rightarrow \infty$. And hence, high order of $q$ could be considered.

\appendices
\section{Proof of Proposition 1}
Here we show that for all $q>2$, a class of polarizing transforms that are equidistant provides the best achievable distance spectrum upper bound. To accomplish this claim, first we prove   
$$2Q\left(\sqrt{\frac{a^2+b^2}{2}}\right) < Q\left(a\right)+Q\left(b\right)$$ 
for $a> 0$, $b>0$ and $a\neq b$.

\begin{figure}[hp]
\centering
\begin{tikzpicture}[thick,scale=1.3, every node/.style={scale=0.99}]
1.a. \draw[<->] (5.2,0)  -- (0,0) --
    (0,5.2);
1. \draw[<->] (5.2,0) node[below]{$t$} -- (0,0) --
    (0,5.2) node[left]{$\psi(t)$};
2. \draw[very thick] (0,5) to [out=270,in=180] (5,0.5);
3. \draw[very thick,dashed] (0.5,3)  -- (4.5,0.5);
4. \draw[very thin,dotted] (-0.1,3) node[left]{$\psi(a)$} -- (0.5,3);
5. \draw[very thin,dotted] (-0.1,0.5) node[left]{$\psi(b)$} -- (4.5,0.5);
6. \draw[very thin,dotted] (0.5,-0.1) node[below]{$a$} -- (0.5,3);
7. \draw[very thin,dotted] (4.5,-0.1) node[below]{$b$} -- (4.5,0.5);
8. \draw[very thin,dotted] (2.5,-0.1) node[below]{$\frac{a+b}{2}$} -- (2.5,1.75);
9. \draw[very thin,dotted] (-0.1,1.75) node[left]{$\frac{\psi(a)+\psi(b)}{2}$} -- (2.5,1.75);
10. \draw[very thin,dotted] (-0.1,1.1) node[left]{$\psi\left(\frac{a+b}{2}\right)$} -- (2.5,1.1);
\end{tikzpicture}
\caption{A graphical interpretation of Jensen's inequality.}
\label{fig_graph_rep}
\end{figure}
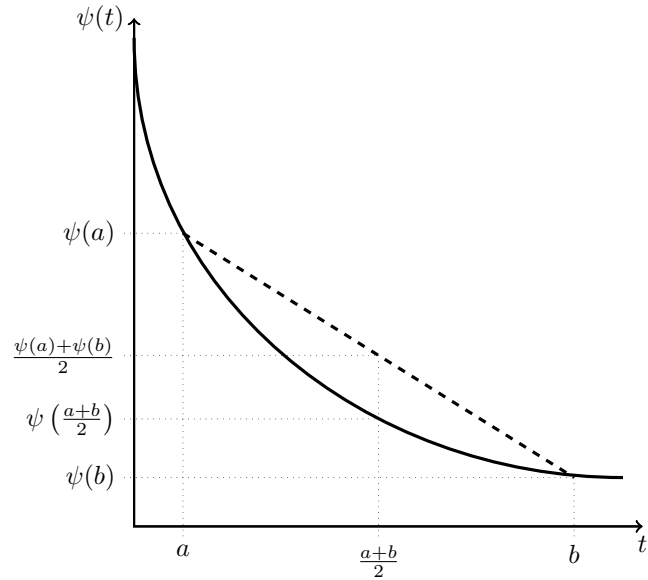

By the use of Jensen's inequality \cite{jensens} for a decreasing concave function $\psi$, following result is obtained.
$$\psi\left(\frac{a+b}{2}\right)<\frac{\psi(a)+\psi(b)}{2}$$
A graphical interpretation of this fact is depicted in Fig.~\ref{fig_graph_rep}.

The proof is completed by the following step.
For all $a>0$, $b>0$
$$\sqrt{\frac{a^2+b^2}{2}}>\frac{a+b}{2}.$$ 
Hence, it is obvious that
$$Q\left(\sqrt{\frac{a^2+b^2}{2}}\right)<Q\left(\frac{a+b}{2}\right)<\frac{Q(a)+Q(b)}{2}.$$
Finally, this result can be generalized for any $q>2$ by using multiple $q-1$ distance terms instead of $a$ and $b$.

\section{Distance spectrum upper bound of the almost-equidistant transform}
Hence, the distance spectrum upper bound of the almost-equidistant transform is
$$P_{e}\leq6Q\left(2\sqrt{\frac{SNR}{2}}\right)+Q\left(2.83\sqrt{\frac{SNR}{2}}\right).$$
If an equidistant transform existed for 8-PSK signal set, the minimimum distance would be $2.14\sqrt{E_s}$, and the upper bound would be  
$$P_{e}\leq7Q\left(2.14\sqrt{\frac{SNR}{2}}\right).$$

\begin{figure}[t!]
\centering
\includegraphics[width=0.6\textwidth]{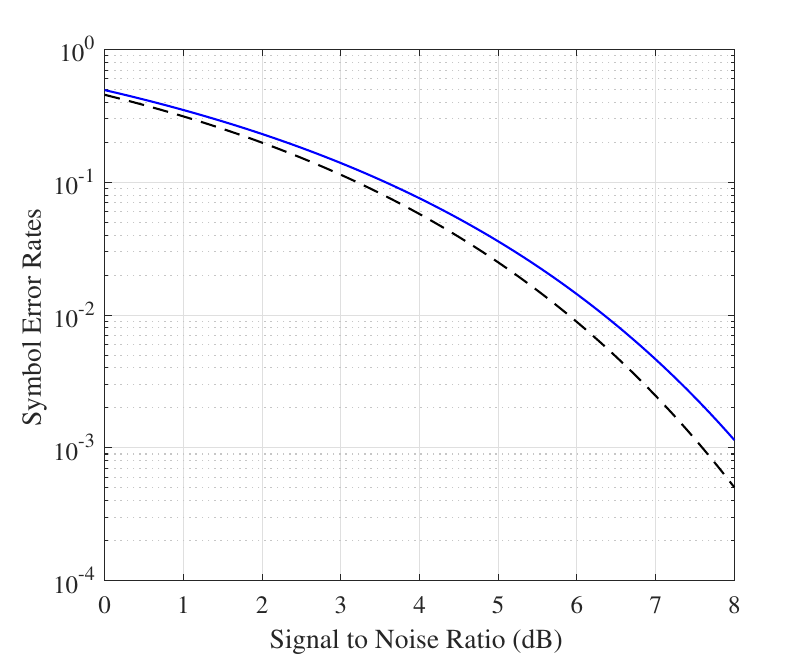}
\caption{Performance of almost-equidistant transform.}
\label{fig_almost_eqd}
\end{figure}

\section*{Acknowledgment}

This work was performed in postdoc study at Bilkent University in Nov. 2015 - Nov. 2017 and the short visit at The Hong Kong Polytechnic University in Dec. 2018 - Jan. 2019. During the postdoc study author was supported by the Scientific and Technological Research Council of Turkey (T\"{U}B\.ITAK), grant: 1929B011500065. 
This work was partly submitted and accepted at ICSEE 2018 in Eilat Israel.
Author would like to thank Prof. Emanuele Viterbo (Monash University), Prof. Ram Zamir (Tel Aviv University) and Prof. Erdal Ar\i kan (Bilkent University) for helpful communications. 

\ifCLASSOPTIONcaptionsoff
  \newpage
\fi

\bibliography{q-ary_Ref}{}
\bibliographystyle{IEEEtran}

\end{document}